%% file: ms.tex
\documentclass[apj,iop,graphicx]{emulateapj}
\usepackage{apjfonts}

\newcommand\swi{{\it Swift}}

\makeatletter
\newcommand*{\rom}[1]{\expandafter\@slowromancap\romannumeral #1@}

\begin{document}

\title{Accretion Disk Reverberation with {\it Hubble Space Telescope} Observations of NGC 4593: Evidence for Diffuse Continuum Lags}
\shortauthors{Cackett et al.}
\shorttitle{NGC 4593 with {\it HST}}

\author{Edward~M.~Cackett\altaffilmark{1}}
\author{Chia-Ying~Chiang\altaffilmark{1}}
\author{Ian~McHardy\altaffilmark{2}}
\author{Rick~Edelson\altaffilmark{3}}
\author{Michael~R.~Goad\altaffilmark{4}}
\author{Keith~Horne\altaffilmark{5}}
\author{Kirk~T.~Korista\altaffilmark{6}}

\email{ecackett@wayne.edu}

\affil{\altaffilmark{1}Department of Physics \& Astronomy, Wayne State University, 666 W. Hancock St, Detroit, MI 48201, USA}
\affil{\altaffilmark{2}University of Southampton, Highfield, Southampton, SO17 1BJ, UK }
\affil{\altaffilmark{3}University of Maryland, Department of Astronomy, College Park, MD 20742-2421, USA}
\affil{\altaffilmark{4}University of Leicester, Department of Physics and Astronomy, Leicester, LE1 7RH, UK}
\affil{\altaffilmark{5}SUPA Physics and Astronomy, University of St. Andrews, Fife, KY16 9SS Scotland, UK}
\affil{\altaffilmark{6}Department of Physics, Western Michigan University, 1120 Everett Tower, Kalamazoo, MI 49008-5252, USA}

\begin{abstract} 
The Seyfert 1 galaxy NGC 4593 was monitored spectroscopically with the {\it Hubble Space Telescope} as part of a reverberation mapping campaign that also included {\it Swift}, {\it Kepler} and ground-based photometric monitoring. During 2016 July 12 -- August 6, we obtained 26 spectra across a nearly continuous wavelength range of $\sim$1150 -- 10,000\AA.  These were combined with {\it Swift} data to produce a UV/optical ``lag spectrum'', which shows the interband lag relative to the {\it Swift} UVW2 band as a function of wavelength.  The broad shape of the lag spectrum appears to follow the $ \tau \propto \lambda^{4/3}$ relation seen previously in photometric interband lag measurements of other active galactic nuclei (AGN).  This shape is consistent with the standard thin disk model but the magnitude of the lags implies a disk that is a factor of $\sim$3 larger than predicted, again consistent with what has been previously seen in other AGN.  In all cases these large disk sizes, which are also implied by independent gravitational microlensing of higher-mass AGN, cannot be simply reconciled with the standard model.  However the most striking feature in this higher resolution lag spectrum is a clear excess around the 3646\AA\ Balmer jump.  This strongly suggests that diffuse emission from gas in the much larger broad-line region (BLR) must also contribute significantly to the interband lags.  While the relative contributions of the disk and BLR cannot be uniquely determined in these initial measurements, it is clear that both will need to be considered in comprehensively modeling and understanding AGN lag spectra.
\end{abstract}

\section{Introduction} \label{sec:intro}

Reverberation mapping \citep{blandmckee82,peterson14} allows an estimate of the size scale of the broad line region (BLR) in Seyfert 1s, and has led to the measurement of approximately 60 black hole masses \citep[e.g.,][]{petersonetal04,bentz09, bentzkatz15}.  The concept of reverberation mapping is straightforward; the observed time lag, $\tau$, between an emission line lightcurve and the optical continuum lightcurve is interpreted as the light-travel time from the continuum emitting region close to the black hole and the line-emitting region further out (assuming that the optical continuum is a good proxy for the driving ionizing continuum).  The emissivity-weighted average radius of the BLR, $R$, is therefore related to the lag via $R = \tau c$.  Assuming that the gas in the BLR is virialized, combining the velocity dispersion of the emission line leads to a black hole mass estimate.

Reverberation mapping can go beyond these very simple mass estimates.  Use of changes in the velocity profile of emission lines can allow the structure of the BLR to be mapped \citep[e.g.][]{welsh91,hornepcn04}.  Recent developments in modeling \citep{brewer11,pancoast14} along with better data have started to improve our understanding of the structure of the BLR, especially the recent AGN STORM campaign to monitor NGC~5548 \citep{derosa15,edelson15,fausnaugh16,goad16,pei17, starkey17,mathur17}. 

While emission line variability probes the BLR, continuum studies allow us to study the accretion disk \citep[e.g.,][]{collier99,cackett07}. If a central source of X-ray/EUV photons irradiates the accretion disk, then correlated continuum variability with wavelength-dependent lags are expected.  In such a scenario, the accretion disk reprocesses high energy EUV/X-ray photons from the central engine into UV/optical continuum photons, with the hot inner regions emitting mainly UV photons and the cool outer regions emitting mainly optical photons.  Thermal radiation from a disk annulus at temperature $T(R)$ emerges with a range of wavelengths, $\lambda \sim hc/kT(R)$. Roughly speaking, each wavelength picks out a different temperature zone and the time lag between the continuum at different wavelengths $\tau = R/c$ measures the corresponding radius.  Thus, shorter wavelengths sense disk annuli at higher temperatures.

More specifically, the observed lags between different continuum wavelengths depend on the disk's radial temperature distribution $T(R)$, which in turn depends on the accretion rate, $\dot{M}$, and the black hole mass, $M$. A disk surface with $T \propto R^{-b}$ will reverberate with a lag spectrum $\tau \propto \lambda^{-1/b}$. For the temperature distribution of a steady-state externally irradiated disk, $T(R)\propto(M\,\dot{M})^{1/4}R^{-3/4}$, the wavelength-dependent continuum lags should follow \citep[see][for more details]{cackett07}:
\begin{equation}
 \tau = \frac{R}{c} \propto (M\dot{M})^{1/3}T^{-4/3}
 \propto (M\dot{M})^{1/3}\lambda^{4/3}  \, . 
\end{equation}

For two decades, hints of wavelength-dependent lags have been observed in a number of AGN \citep{wanders97,collier98,collier01,sergeev05,cackett07,breedt09}, and show an increase in lag with wavelength, with the lags being of the order of a few days for the most massive objects.  However, the data have often suffered from poor temporal sampling (given the short lags) leading to large uncertainties in the lag.  Several recent observing campaigns have changed this \citep{mchardy14, shappee14,edelson15,fausnaugh16,edelson17}.  In particular,  the AGN STORM campaign on NGC~5548 used a combination of monitoring with the {\it Hubble Space Telescope} ({\it HST}) and the {\it Neil Gehrels Swift Observatory} ({\it Swift} hereafter), obtaining 282 observations over 125 days with a mean sampling rate of less than 0.5 days \citep{edelson15}.  Moreover, they used all 6 {\it Swift}/UVOT filters for the first time in a large AGN monitoring campaign.  \citet{fausnaugh16} also include ground-based monitoring in 9 additional bands during that campaign, leading to the best characterized wavelength-dependent lags to date.

The wavelength-dependent lags in NGC~5548, however, challenge our understanding of AGN disks.  While the measured lags appear to generally follow the predicted $\tau \propto \lambda^{4/3}$ relation (though the best-fitting slope is slightly flatter), the lags appear larger than expected at all wavelengths.   If one makes the standard assumption that $L/L_{\rm Edd} = 0.1$ then the disk is approximately a factor of 3 larger than the standard prediction \citep{mchardy14,edelson15,fausnaugh16}.  Also interesting is that the lags for the $u$ and $U$ bands appear slightly increased compared to the general wavelength-dependent trend \citep{edelson15, fausnaugh16}. 

One possibility is that the continuum lags are affected by more slowly varying components arising in the BLR. For example, \citet{koristagoad01} showed that continuous radiation emanating from dense ($n_H > 10^{10}$~cm$^{-3}$) BLR clouds is significant, and lags the driving continuum in a manner that generally increases with increasing wavelength across the UV to near-IR.  In particular, it produces a substantially enhanced delay shortward of the Balmer jump. This diffuse continuum is comprised of thermal free-bound and free-free continua plus scattered incident continuum from BLR clouds, which will respond to ionizing continuum variations on timescales longer than the light-travel timescale to the accretion disk, but generally substantially shorter than those associated with the gas emitting Ly$\alpha$ \citep{koristagoad01}.  The much higher gas densities and relatively higher electron temperatures found in BLR clouds account for the enhanced strength of the diffuse continuum, relative to that emanating from the narrow line region or \ion{H}{2} regions.  The strength of the diffuse continuum component is sensitive to the presence of high gas densities and ionizing photon fluxes, which make it an important diagnostic of the physical conditions within the BLR.  Since this diffuse continuum's contribution will act to lengthen the measured lags above that from the accretion disk, it is important to try to assess its contribution.

The AGN STORM NGC~5548 campaign raised another important question -- the role that the X-rays play in driving the variability in the UV/optical lightcurves.  MCMC modeling of the lightcurves from this campaign by \citet{starkey17} show that reprocessing in the accretion disk is consistent with the UV/optical lightcurves.  However, the driving lightcurve that is recovered from this fitting does not match the observed X-ray lightcurve \citep{starkey17}.  \citet{gardnerdone17} also show that blurring the hard X-ray lightcurve gives too much fast variability, and suggest that there is an intervening puffed-up Comptonized disk region that blocks the X-rays from illuminating the disk directly.  

In 2016, a high-cadence {\it Swift} monitoring campaign of the bright AGN NGC 4151 took place \citep{edelson17}, with approximately 6 hour sampling over a 69-day period.  Here, the X-ray/UV/optical lightcurves are all well correlated.  However, the UV/optical lightcurves lag the X-ray lightcurve by about 3 -- 4 days, while the UV to optical lags are less than 1 day.  This disconnect between the X-ray to UV and UV to optical lags is even more strongly in conflict with the standard model.  One possible explanation could be that an additional component shields the disk from the corona and reprocesses the energy on a longer (e.g. dynamical) timescale, increasing the lag between the X-rays and optical/UV \citep{gardnerdone17,edelson17}.

The AGN STORM campaign on NGC~5548 is not the only study to imply that AGN accretion disks are too large.  From fitting accretion disk models to the wavelength-dependent lags and observed fluxes from \citet{sergeev05}, \citet{cackett07} determined a value of $H_{\rm 0}$ a factor of 1.6 too small.  This is the equivalent to the disks being too big (the lag too large) by a factor of 1.6 based on the observed flux compared to the accretion disk model.  More recently, observations of NGC~2617 \citep{shappee14}, NGC~3516 \citep{noda16}, NGC~6814 \citep{troyer16}, Fairall~9 \citep{pal16}, Ark~120 \citep{gliozzi17}, and a sample of 21 AGN in the {\it Swift} archive \citep{buisson17}, all find lags that are longer than expected for a standard thin disk.  Both Pan-STARRS and the Dark Energy Survey (DES) are obtaining lightcurves of quasars in multiple photometric bands, allowing for determining the average sizes from a large number of objects \citep{jiang17,mudd17}.  From Pan-STARRS lightcurves of 240 quasars, \citet{jiang17} conclude that the lags are $\sim$2 -- 3 times larger than for a standard thin disk. \citet{mudd17} model lightcurves from 15 DES quasars, finding they can be well fit by a thin disk model if they are accreting at moderate Eddington rates ($\sim$0.3).  A completely independent method of determining accretion disk sizes using gravitational microlensing, also finds disks are larger than expected by the standard picture \citep[e.g.,][]{morgan10,dai10,mosquera13}.  

A common picture seems to be arising, then, where accretion disks appear to be larger than predicted.  In order to test this, and to better understand the possible contribution to the lags from the diffuse continuum emission, we undertook a multi-wavelength campaign on NGC~4593 during July 2016.  NGC~4593 is a nearby ($z = 0.0087$), highly variable Seyfert~1 which has shown significant broad emission line lags in H$\beta$ \citep{denney06,barth13} as well as H$\gamma$ and \ion{Fe}{2} \citep{barth13}.  It has a black hole mass estimated by reverberation mapping to be $M = (7.63\pm1.62)\times10^6~M_\odot$ \citep{bentzkatz15}.

Our campaign took advantage of the fact that NGC~4593 was in the field of view of the {\it Kepler} satellite at the time, and we coordinated observations with {\it Swift}, {\it HST} and ground-based observatories to coincide with this.  Here, we present the {\it HST} observations from this campaign, focusing on the continuum lightcurves and time lags.  Analysis of the emission line properties is left to future work.  An accompanying paper on the {\it Swift} lightcurves is presented by \citet{mchardy17}, though see also \citet{palnaik17} who also analyze the same {\it Swift} data.

We briefly summarize the main results of \citet{mchardy17} here.  {\it Swift} performed 194 observations of NGC~4593 over a 22.6 day period, obtaining lightcurves in all 6 UV/optical filters and in X-rays. Timing analysis shows that the optical (B and V) bands lag the UVW2 band by $\sim$0.2 days.  Fitting a $\lambda^{4/3}$ relation reveals that the $U$ band lag is enhanced and the X-ray lag ($\sim$0.7 days) is significantly offset from the best-fitting relation fit. Maximum entropy modeling of the lightcurves reveals that the shape of the X-ray lightcurve is consistent with being the driving lightcurve, and that the response function in the UV/optical bands is consistent with a combination of a strong prompt response in addition to a weaker response on long timescales.  In this paper, we present the {\it HST} observations obtained during the same campaign.

\section{Observations and Data Reduction}\label{sec:data}

\begin{figure}
\centering
\includegraphics[width=\columnwidth]{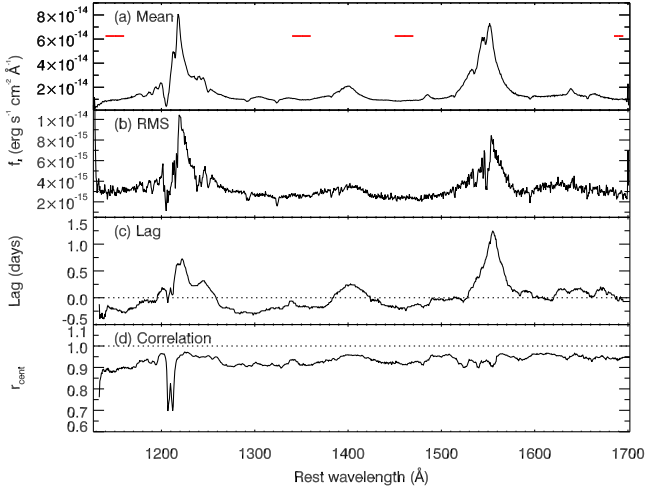}
\caption{(a) Mean spectrum, (b) rms spectrum, (c) lag spectrum measured with respect to the {\it Swift}/W2 band and (d) $r_{\rm cent}$, the correlation coefficient at the lag centroid, from the 26 HST observations for the G140L grating. Red horizontal lines indicate the chosen continuum bands.}
\label{fig:meanrmsg140}
\end{figure}

\begin{figure}
\centering
\includegraphics[width=\columnwidth]{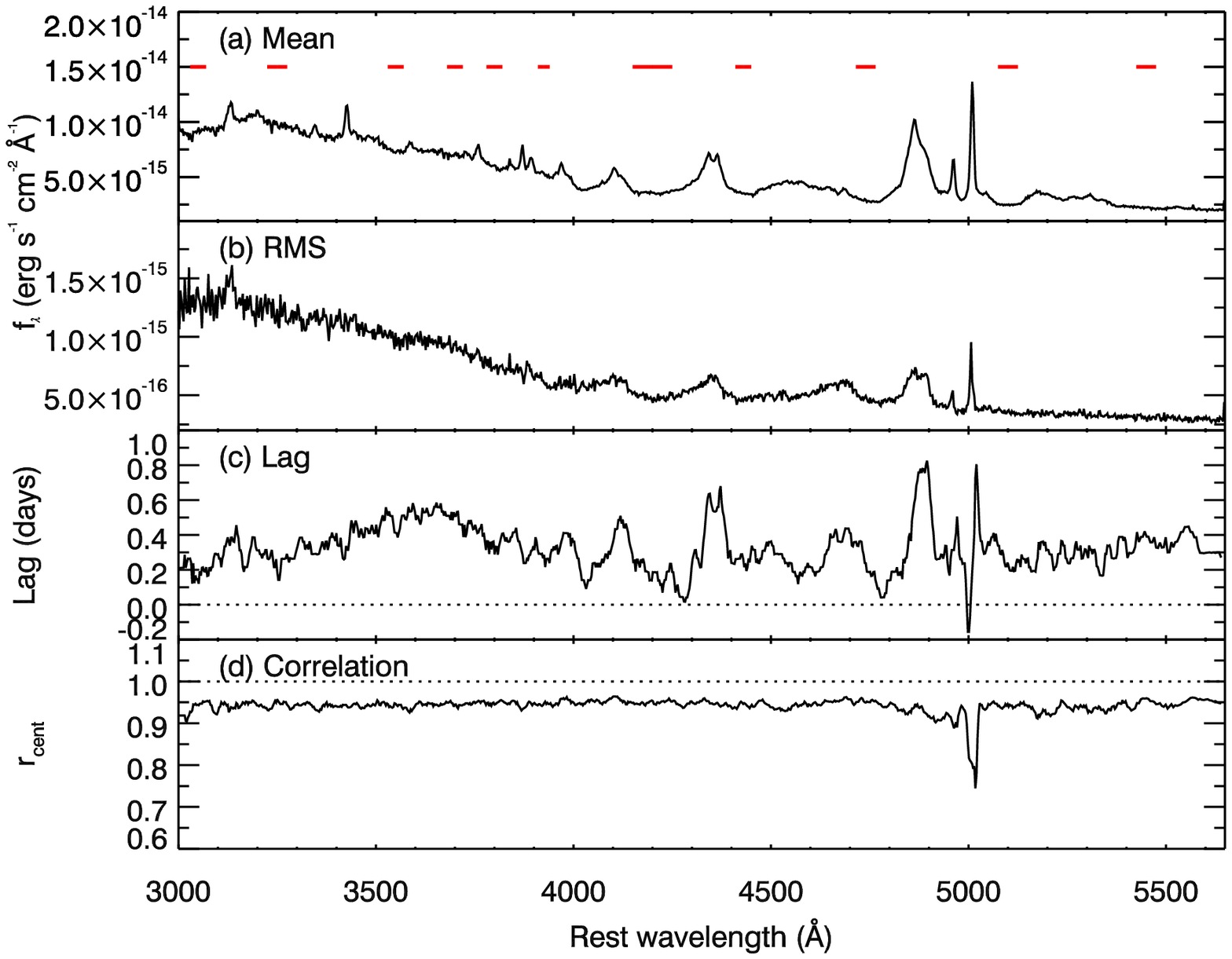}
\caption{(a) Mean spectrum, (b) rms spectrum, (c) lag spectrum measured with respect to the {\it Swift}/W2 band and (d) $r_{\rm cent}$, the correlation coefficient at the lag centroid, from the 26 HST observations for the G430L grating. Red horizontal lines indicate the chosen continuum bands.}
\label{fig:meanrmsg430}
\end{figure}

\begin{figure}
\centering
\includegraphics[width=\columnwidth]{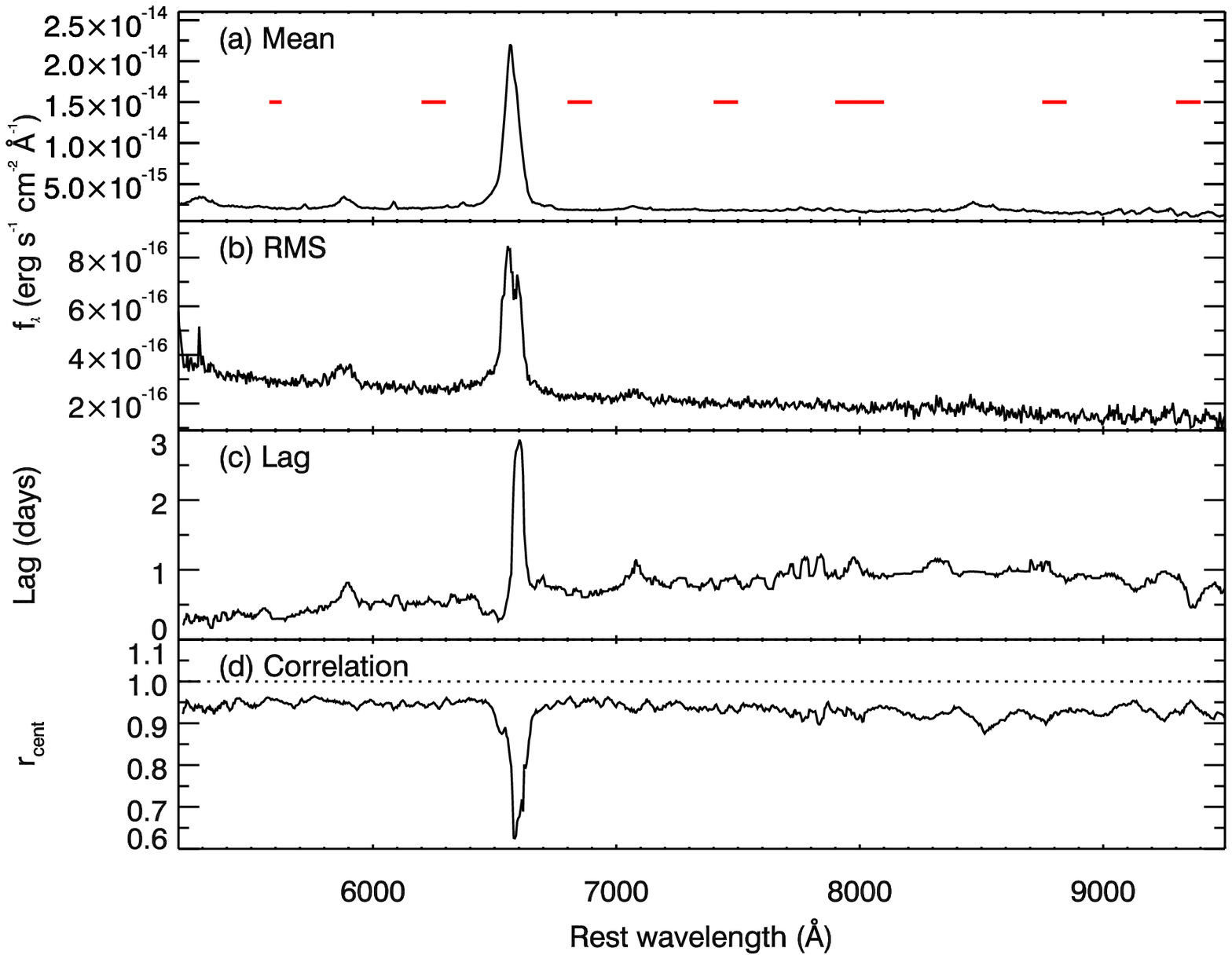}
\caption{(a) Mean spectrum, (b) rms spectrum, (c) lag spectrum measured with respect to the {\it Swift}/W2 band  and (d) $r_{\rm cent}$, the correlation coefficient at the lag centroid, from the 26 HST observations for the G750L grating. Red horizontal lines indicate the chosen continuum bands.}
\label{fig:meanrmsg750}
\end{figure}

\begin{figure}
\centering
\includegraphics[width=\columnwidth]{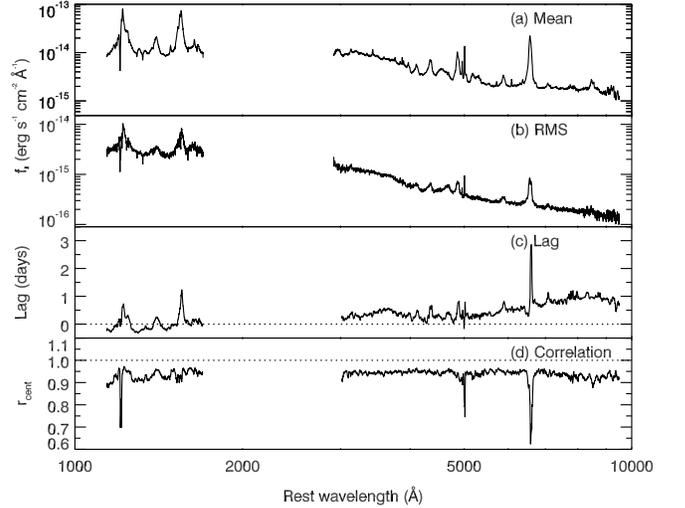}
\caption{(a) Mean spectrum, (b) rms spectrum, (c) lag spectrum measured with respect to the {\it Swift}/W2 band and (d) $r_{\rm cent}$, the correlation coefficient at the lag centroid, from the 26 HST observations for all wavelengths covered. }
\label{fig:meanrms_all}
\end{figure}

Single-orbit observations of NGC~4593 were performed approximately daily by {\it HST} from 2016 July 12 through August 6, with 26 out of 27 scheduled observations successfully executed.  Since the main goal of this campaign is to study the wavelength-dependent continuum lags, the observations were designed to efficiently cover as broad a wavelength range as possible, hence we used the Space Telescope Imaging Spectrograph (STIS) with low-resolution gratings.  During each observation, spectra were obtained with the G140L, G430L and G750L gratings and the $52 \times 0.2$\arcsec\ aperture, using a fixed position angle of 67$^\circ$ E of N.  The fixed PA ensures the same contribution from the host galaxy and any extended narrow line region in each observation.   For the G140L we took a 1234s exposure using a central wavelength of 1425\AA\, giving a wavelength range of 1119--1715\AA.  G430L had a central wavelength of 4300\AA, exposure of 298s and wavelength range 2888--5697\AA.  For the G750L grating, we used a central wavelength 7751\AA, total exposure of 288s and wavelength range 5245--10233\AA.  

The standard pipeline-processed spectra show many streaks on the STIS CCDs that are not removed by the standard processing.  These show up as sharp (one-pixel) spikes in the mean and rms spectra.  We therefore used the \verb|stis_cti| package created by the COS/STIS team to apply Charge Transfer Inefficiency (CTI) corrections to the data. The script uses a pixel-based correction algorithm based on \citet{andersonbedin10} and removes trails caused by CTI effects in the CCDs. These corrections significantly improved the spectra with only a few hot pixels remaining. We remove these small number of defects manually by linearly interpolating the flux from neighboring pixels.

\section{Data Analysis}\label{sec:analy}

\begin{figure*}
\centering
\includegraphics[width=0.8\textwidth]{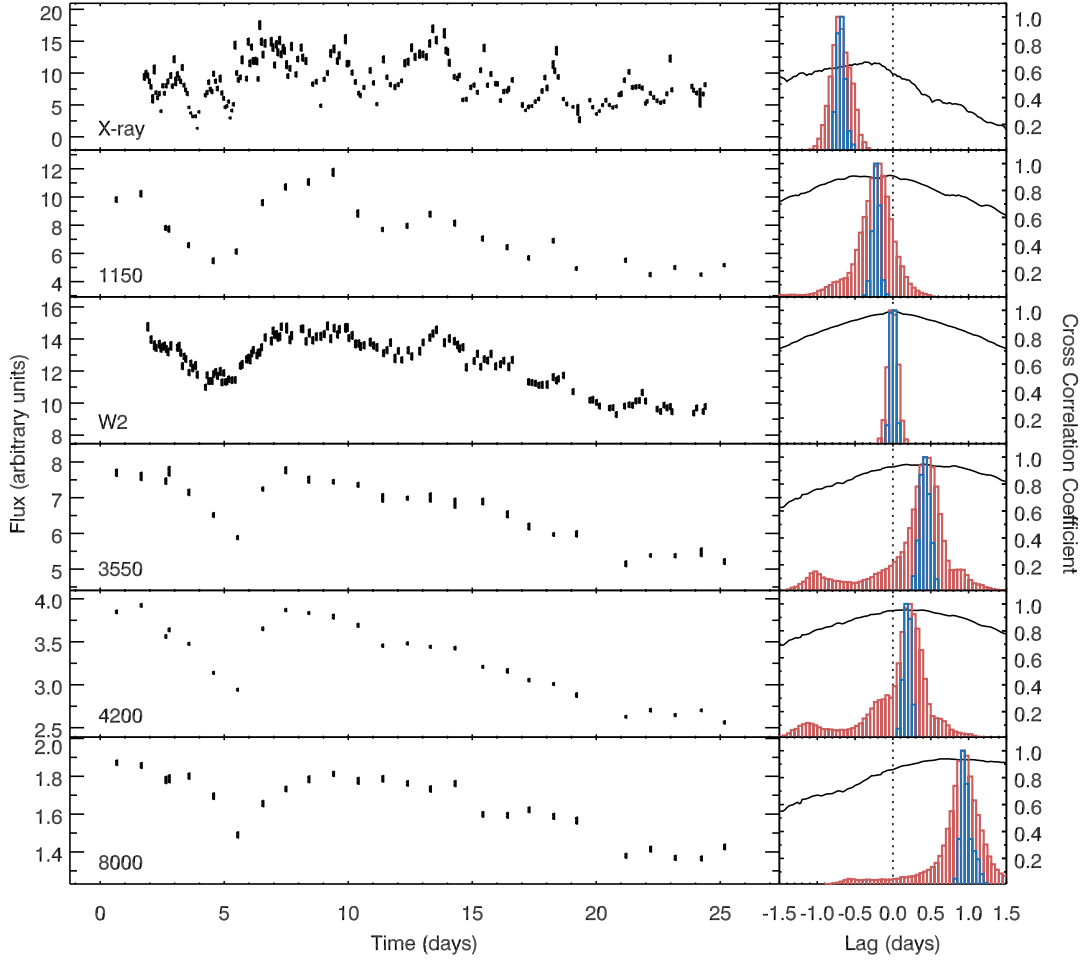}
\caption{{\it Left:} Lightcurves for selected {\it HST} and {\it Swift} bands.  Labels indicate either the {\it Swift} band, or the {\it HST} wavelength (\AA). {\it Right:} Cross-correlation functions (solid black line) and cross-correlation centroid distributions (histograms), calculated with respect to the {\it Swift}/W2 lightcurve for the full FR+RSS uncertainty method (red) and the FR only method (blue).}
\label{fig:lc}
\end{figure*}

\subsection{Mean and rms spectra}

We created mean and rms spectra from the 26 observations in each of the gratings, following the standard definition \citep[e.g., Equations 2 and 3 in][]{petersonetal04}.  All wavelengths and fluxes shown are at their rest frame values (using $z = 0.0087$), and have been dereddened using \citet{cardelli89} and assuming $E(B-V) = 0.021$ \citep{schlafly11}.  We show the mean and rms spectra separately for each grating in Figures \ref{fig:meanrmsg140}, \ref{fig:meanrmsg430} and \ref{fig:meanrmsg750}, and for all wavelengths combined in Figure~\ref{fig:meanrms_all}.  These figures also show the lag spectrum and correlation coefficient at the lag centroid (both described in Section~\ref{sec:lagspec}). The G430L and G750L spectra overlap slightly in wavelength, and for figures including the full wavelength range we show the G430L for $\lambda < $5400\AA.  There is excellent agreement between the mean and rms spectra in the overlapping regions.

The rms spectrum shows significant variability in both the broad emission lines and continuum, with the continuum variability amplitude generally decreasing with increasing wavelength (see variability amplitudes quoted in Table~\ref{tab:contlags}).

\subsection{Continuum lightcurves}

We begin our time series analysis by identifying line-free regions of the spectrum in order to extract clean continuum lightcurves.  We do this initially by selecting 22 wavelength bands spanning the full wavelength range of the spectra.  The wavelength bands vary in width, depending on the presence of emission lines, from 10 \AA\ to 200 \AA.  The chosen continuum bands are shown as red horizontal lines in Figures~\ref{fig:meanrmsg140}, ~\ref{fig:meanrmsg430}, and~\ref{fig:meanrmsg750} and the exact wavelength ranges used are given in Table~\ref{tab:contlags}.  We take the mean flux within each band at each epoch in order to create continuum lightcurves at each wavelength.  We show four of the {\it HST} continuum lightcurves, along with the {\it Swift} X-ray and W2 lightcurves from \citet{mchardy17} in Figure~\ref{fig:lc}.  The {\it HST} lightcurves for each grating are given in the Appendix.

The {\it Swift} lightcurves of NGC~4593 during the monitoring campaign are sampled at a significantly higher cadence than we could obtain with {\it HST}.  We therefore chose the shortest wavelength UV lightcurve from {\it Swift} (the W2 filter) as the reference band for our cross-correlation analysis.  Of the UVOT bands, the W2 lightcurve has the highest variability amplitude and S/N ratio, and thus combined with the improved cadence leads to the best determined lag measurements.  The W2 lightcurve is preferred over the X-ray lightcurve since it gives a significantly higher peak correlation coefficient, and thus better constrained lags. The {\it Swift} data reduction, analysis and lightcurves are described in detail in \citet{mchardy17}, and we use those same lightcurves here.

In order to assess the wavelength-dependent time lags, we perform a cross-correlation analysis with respect to the {\it Swift}/W2 lightcurve following the interpolated cross-correlation function (ICCF) method, as described by \citet{white94}.  A detailed discussion of the uncertainties is given in Section~\ref{sec:uncert}.  

\begin{figure}
\centering
\includegraphics[width=\columnwidth]{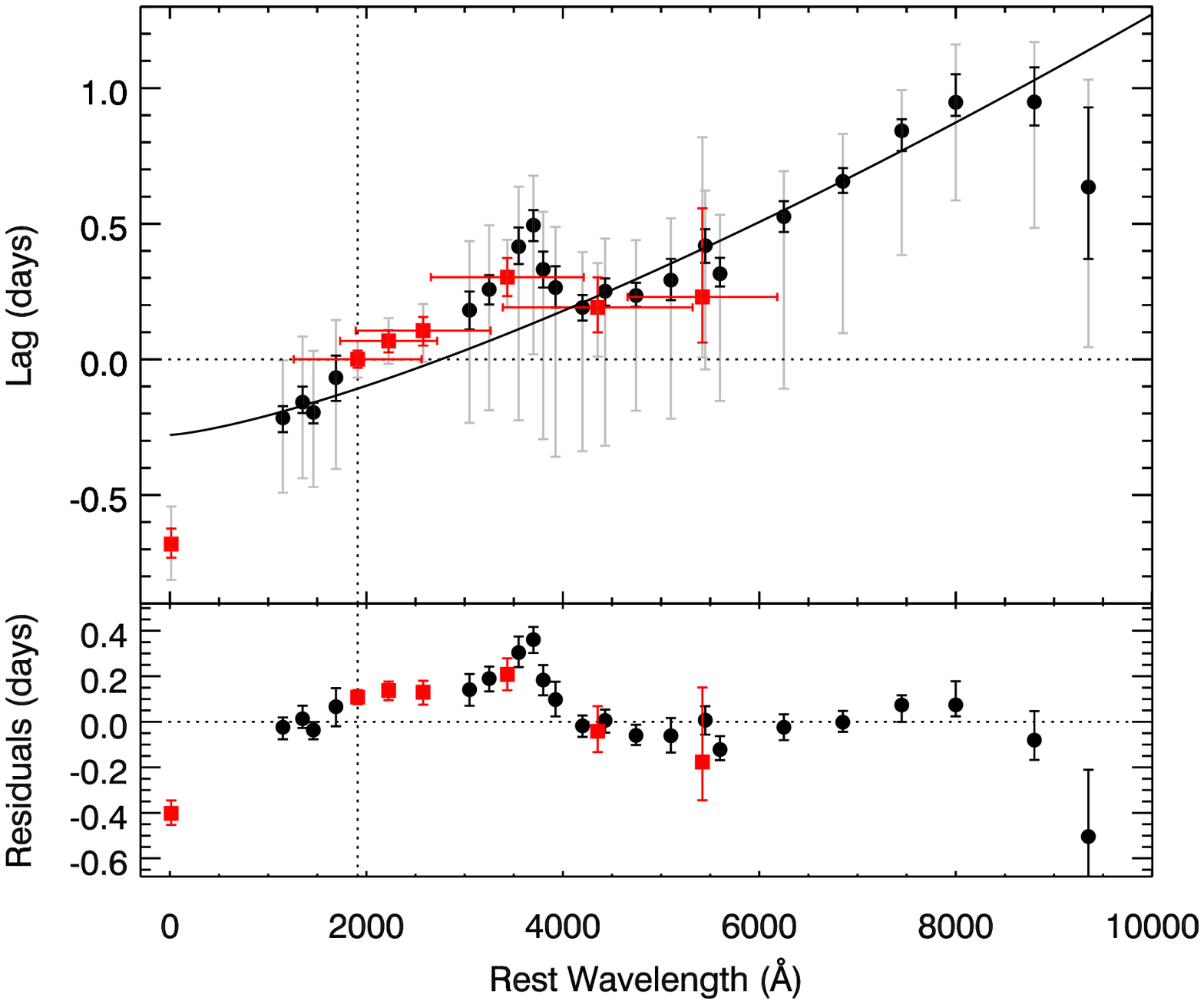}
\caption{Lag as a function of wavelength for {\it HST} (black, circles) and {\it Swift} (red, squares) bands, calculated with respect to the {\it Swift} W2 band (wavelength indicated by a vertical dotted line).  The solid line shows the best-fitting $\tau \propto \lambda^{4/3}$ relation when ignoring the points from 3000 -- 4000\AA.  Black and red uncertainties are calculated from FR only, while gray uncertainties are from the full FR+RSS (see text for details).  A clear excess and discontinuity in the lags is seen around the Balmer jump (3646~\AA), which is highlighted further when looking at the residuals (bottom panel).}
\label{fig:contlags}
\end{figure}

The lags are given in Table~\ref{tab:contlags}, along with the variability amplitude, $F_{\rm var}$, calculated following \citet{vaughan03}, and the correlation coefficient at the centroid lag, $r_{\rm cent}$ (note that $r_{\rm cent}$ is high at all wavelengths).  We also give the lags of the lightcurves from each of the {\it Swift} filters in Table~\ref{tab:swlags}.  The lags are plotted as a function of wavelength in Figure~\ref{fig:contlags}.  An increase in lag with wavelength is observed, except around the Balmer jump (3646\AA), where a clear discontinuity can be seen. The lags are also suggestive of a Paschen jump (8204\AA) which is also predicted by photoionization models for the diffuse continuum, though the uncertainty in the lag is larger there, making the drop not statistically significant  (the lag at 9350\AA\ is less than 2$\sigma$ below the best-fitting $\lambda^{4/3}$ relation).  We discuss the discontinuity around the Balmer jump in Section~\ref{sec:model}.

\begin{deluxetable*}{ccccccc}
\tablecaption{{\it HST} continuum lightcurve variability, and time lags calculated with respect to the {\it Swift} W2 lightcurve. \label{tab:contlags}
}
\tablehead{Wavelength (\AA) & Range (\AA) & $F_{\rm var}$ & $r_{\rm cent}$ & Lag (days) & \multicolumn{2}{c}{Lag Uncertainty (days)}\\
 & & & & & FR only & FR+RSS}
\startdata
1150 & 1140 -- 1160 & $0.286\pm0.005$ &  0.89 & $-0.22$ &    $-0.05,  +0.04$ &   $-0.28, +0.21$ \\
1350 & 1340 -- 1360 & $0.262\pm0.003 $ & 0.91  & $-0.16$ & $-0.04, +0.06$, & $-0.28, +0.24$ \\ 
1460 & 1450 -- 1470 & $0.263\pm0.003$ & 0.92  & $-0.20$ & $\pm0.04$ & $-0.27, +0.23$\\  
1690 & 1685 -- 1695 & $0.241\pm0.006$ & 0.93 & $-0.07$ &  $-0.09, +0.08$ & $-0.34, +0.21$\\ 
3050 & 3030 -- 3070 & $0.133\pm0.003$ & 0.95 & $0.18$ &  $\pm0.07$ &  $-0.42, +0.25$\\ 
3250 & 3225 -- 3275 & $0.120\pm0.002$ & 0.95 &  $0.26$ &  $-0.06, +0.05$ &  $-0.45,  +0.23$\\
3550 & 3540 -- 3560 & $0.129\pm0.003$ & 0.95 &  $0.42$ &  $\pm0.07$ & $-0.64, +0.22$\\ 
3700 & 3680 -- 3720 & $0.129\pm0.002$ & 0.95 &  $0.49$ &  $\pm0.06$ & $-0.48, +0.18$\\
3800 & 3780 -- 3820 & $0.122\pm0.003$ & 0.95 &  $0.33$ & $\pm0.07$ & $-0.63, +0.21$\\ 
3925 & 3910 -- 3940 & $0.113\pm0.003$ & 0.95 &   $0.27$ &  $\pm0.08$ &  $-0.61, +0.22$\\
4200 & 4150 -- 4250 & $0.132\pm0.001$ & 0.95 &   $0.19$ &  $\pm0.05$ & $-0.53, +0.20$\\
4430 & 4410 -- 4450 & $0.128\pm0.002$ & 0.95 &   $0.25$ & $\pm0.05$ &  $-0.57, +0.19$\\
4745 & 4720 -- 4770 & $0.159\pm0.002$ & 0.96 &  $0.24$ & $-0.04, +0.05$ & $-0.42, +0.20$\\  
5100 & 5050 -- 5150 & $0.130\pm0.003$ & 0.95 &  $0.29$ &  $-0.07, +0.08$ & $-0.51, +0.23$\\  
5450 & 5425 -- 5475 & $0.136\pm0.002$ & 0.95 &  $0.42$ &  $\pm0.06$ &  $-0.45, +0.21$\\
5600 & 5575 -- 5625 & $0.137\pm0.002$ & 0.96 &  $0.32$ &  $-0.05, +0.06$ &  $-0.47, +0.22$\\
6250 & 6200 -- 6300 & $0.118\pm0.002$ & 0.95  &  $0.53$ &  $-0.05, +0.06$ &  $-0.64, +0.17$\\
6850 & 6800 -- 6900 & $0.120\pm0.001$ & 0.95 &  $0.66$  & $-0.04, +0.05$ & $-0.57, +0.18$\\  
7450 & 7400 -- 7500 & $0.105\pm0.002$ & 0.94 &  $0.84$ &  $-0.08, +0.04$ & $-0.45,  +0.15$\\ 
8000 & 7900 -- 8100 & $0.097\pm0.002$ & 0.93 & $0.95$ &  $-0.05,  +0.10$ & $-0.36, +0.22$\\   
8800 & 8750 -- 8850 & $0.083\pm0.003$ & 0.92 & $0.95$ &   $-0.09,  +0.13$ & $-0.46, +0.22$\\ 
9350 & 9300 -- 9400 & $0.075\pm0.007$ & 0.95  & $0.63$ &   $-0.26, +0.29$ &  $-0.58,  +0.41$
\enddata
\end{deluxetable*}

\begin{deluxetable}{rcccc}
\tablecaption{Lags from {\it Swift} lightcurves with respect to W2. \label{tab:swlags}
}
\tablehead{Filter & $r_{\rm cent}$ & Lag (days) & \multicolumn{2}{c}{Lag Uncertainty (days)}\\
 & & & FR only & FR+RSS}
\startdata
0.5 -- 10 keV & 0.63 & $-0.68$ &  $-0.05, +0,06$  &   $-0.13, +0.14$\\
W2 & 1.00 & $0.00$ &    $\pm0.03$ & $\pm0.07$\\
M2 & 0.96 & $0.07$ &  $\pm0.04$  & $\pm0.08$ \\
W1 & 0.95 & $0.11$ &   $\pm0.06$ & $-0.12, +0.10$ \\
U &   0.93 & $0.30$ & $\pm0.07$ & $-0.11, +0.14$\\
B &  0.82 & $0.19$ & $-0.09, +0.11$ &  $-0.18, +0.16$\\
V & 0.66 & $0.23$ &  $-0.17, +0.33$ &  $-0.22, +0.59$
\enddata
\end{deluxetable}

\subsection{Lag uncertainties}\label{sec:uncert}

The uncertainties on the measured lags for the ICCF method are usually performed following the flux randomization (FR) and random subset sampling (RSS) approach \citep[as implemented by][]{petersonetal04}.  In that method many realizations of the lightcurves are generated.  For a lightcurve with $N$ data points, a subset is chosen by randomly selecting data points from the lightcurve $N$ times, with replacement.  This results in some points being selected multiple times and others not being selected at all.  The error bar on each flux measurement is scaled by the inverse square root of the number of times that data point has been selected.  After the random subset sampling, the flux of each point is randomized using a Gaussian-distributed random number with mean equal to the observed flux and standard deviation equal to the adjusted error bar.  For each generated pair of lightcurves a new CCF is calculated, and the centroid value determined.  The process is repeated a large number of times to build up a distribution of centroid values from which the median and 1$\sigma$ confidence intervals can be determined.  This method was designed in order to assess both the uncertainty in the lag measurement due to the uncertainty in the flux measurement at each epoch, and also to include the uncertainty in the lag measurement due to the exact sampling of the lightcurve -- the RSS tests the influence of individual epochs on the lag that is measured.

Because the {\it HST} dataset has 22 continuum lightcurves all with the same time sampling, we extend the usual FR+RSS method in order to assess separately the statistical errors due to uncertain flux measurements and systematic errors due to the time sampling.  The specific time sampling of the {\it HST} lightcurves should affect the CCF lag in a similar way at all 22 wavelengths. The RSS step simulates this systematic error.  The flux measurement uncertainties are independent at each wavelength, and the FR step simulates these statistical errors.

\begin{figure}
\centering
\includegraphics[width=\columnwidth]{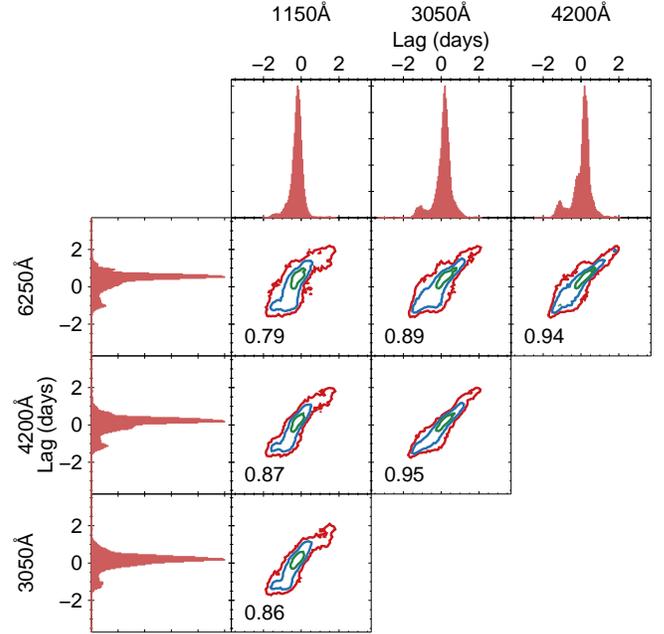}
\caption{Contour plots show the 1$\sigma$ (green), 2$\sigma$ (blue) and 3$\sigma$ (red) 2-dimensional lag centroid distributions from $10^5$ realizations of the lightcurves. Histograms show the 1-dimensional lag centroid distributions.  Numbers in the contour plots show the correlation coefficient, indicating that lag measurements from each realization are highly correlated.}
\label{fig:2dccfd}
\end{figure}

In other words, the lag uncertainties from the RSS stage will be highly correlated between the {\it HST} lightcurves.  We see this when performing the standard FR+RSS technique measuring the lags between the {\it HST} bands and the {\it Swift}/W2 lightcurve.  The point-to-point scatter between lag measurements from neighboring wavelengths is significantly smaller than the size of the uncertainties determined.  To investigate this further, instead of running independent FR+RSS simulations for each {\it HST} wavelength, we use the same RSS for all 22 lightcurves, and then apply FR to each flux measurement.  Thus, for each realization of the sampling we obtain a lightcurve at each wavelength and so can investigate the effect of each sampling on the measured lags, and any correlations between them.  We perform $10^5$ FR+RSS realizations in this way and then look at the lag centroid distributions.  To look for correlations between the bands we look at the two-dimensional lag centroid distributions for each combination of the 22 wavelengths with all the other bands.  We find that they are generally very highly correlated, with lower correlation at longer wavelengths where the FR becomes more important because of the lower variability amplitude and lower S/N there.  We show a subset of the 2D lag centroid distributions in Figure~\ref{fig:2dccfd}.  The contours show the 1, 2 and 3$\sigma$ confidence levels for the 2D distribution.  The contours are highly diagonal, demonstrating that when a higher lag is measured in one waveband, a higher lag is also measured in the other.  The numbers in each box indicate the Pearson's correlation coefficient, again indicating that the distributions are highly correlated between wavelengths.

\begin{figure}
\centering
\includegraphics[width=\columnwidth]{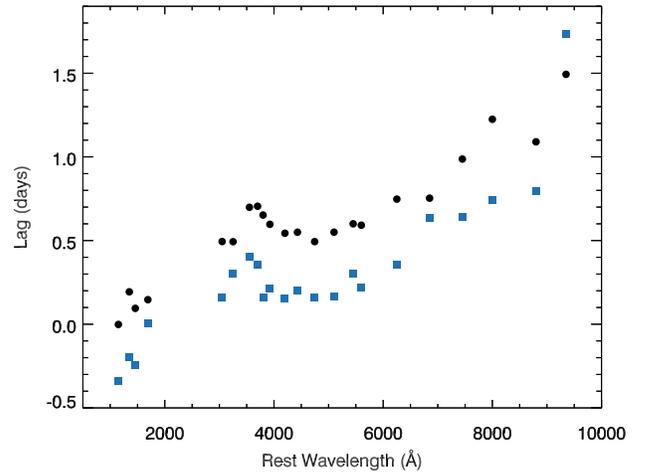}
\caption{Two  realizations chosen to illustrate the systematic correlated effect on the lags from resampling the lightcurves.  Often the lags are generally just shifted up or down with different realizations.}
\label{fig:realizations}
\end{figure}

We also looked at the centroid lags as a function of wavelength determined from individual realizations.  Generally, the lags shift up and down maintaining approximately the same overall wavelength dependence and shape.  We demonstrate this in Figure~\ref{fig:realizations} where we show two realizations towards the upper and lower end of the range of lags.  We note that this general up and down shifting is mostly, but not always, seen and for some realizations the trend with wavelength is more significantly altered.

We check that the correlated errors are caused by the RSS by re-running the simulations using FR only.  When performing FR only we obtain 2D centroid distributions that are not correlated.  Since we are most concerned about the wavelength-dependence of the lags, the systematic offset in lag is less important here.  We care mostly about the trend with wavelength.  We therefore use the FR only lag uncertainties when discussing the wavelength-dependence of the lags. 

Tables~\ref{tab:contlags} and \ref{tab:swlags} give both the FR only and FR+RSS uncertainties for the {\it HST} and {\it Swift} lightcurves.  Note the much smaller difference between the FR only and FR+RSS for the {\it Swift} data since those lightcurves have a much higher temporal sampling.  Next to each lightcurve in Figure~\ref{fig:lc} we plot the corresponding cross-correlation functions, along with a histogram of the lag centroid distribution from both the FR only and FR+RSS simulations.

\subsection{Lag spectrum}\label{sec:lagspec}

In order to explore the lags further we calculate a `lag spectrum' for each grating.  The concept of a lag spectrum has only occasionally been applied to UV/optical spectra of AGN \citep[see][]{collier98,collier99}. However, in recent years it has become common in X-ray reverberation studies to look at the time lag as a function of all energies (see the review on X-ray reverberation by \citealt{uttley14} and a recent example of X-ray lag spectra in \citealt{kara16}).  Applying the same concept here, we want to calculate the time lag (with respect to the reference W2 band) at all wavelengths bins in all the gratings.

\begin{figure}
\includegraphics[width=\columnwidth]{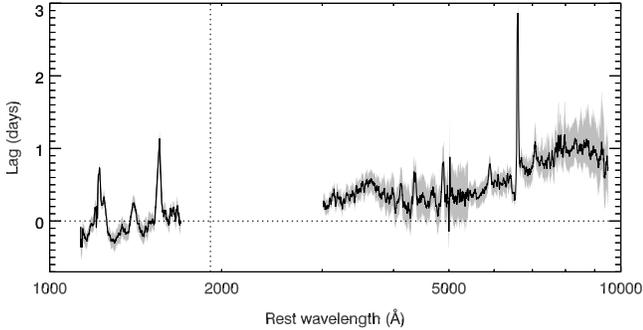}
\caption{Lag spectrum (solid black line) with respect to {\it Swift}/W2 (wavelength indicated by vertical dotted line).  the gray region shows the 1$\sigma$ FR only uncertainties.}
\label{fig:lagspec}
\end{figure}

\begin{figure}
\includegraphics[width=\columnwidth]{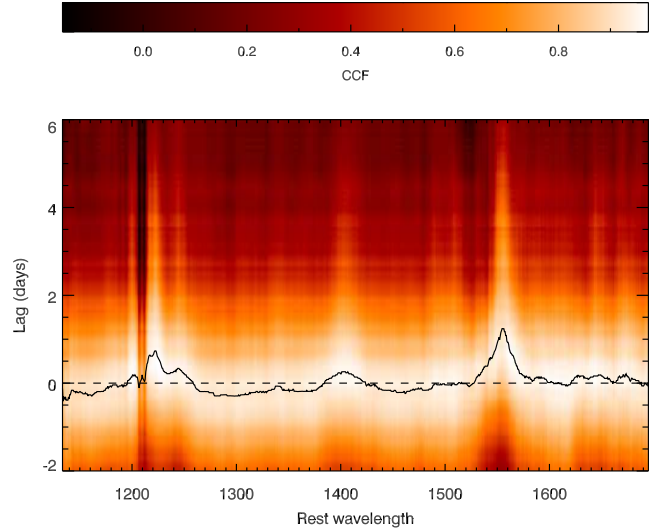}
\caption{Color map shows the CCF (with respect to {\it Swift}/W2) for the G140L grating.  The black line indicates the centroid of the CCF.}
\label{fig:ccfg140}
\end{figure}

\begin{figure}
\includegraphics[width=\columnwidth]{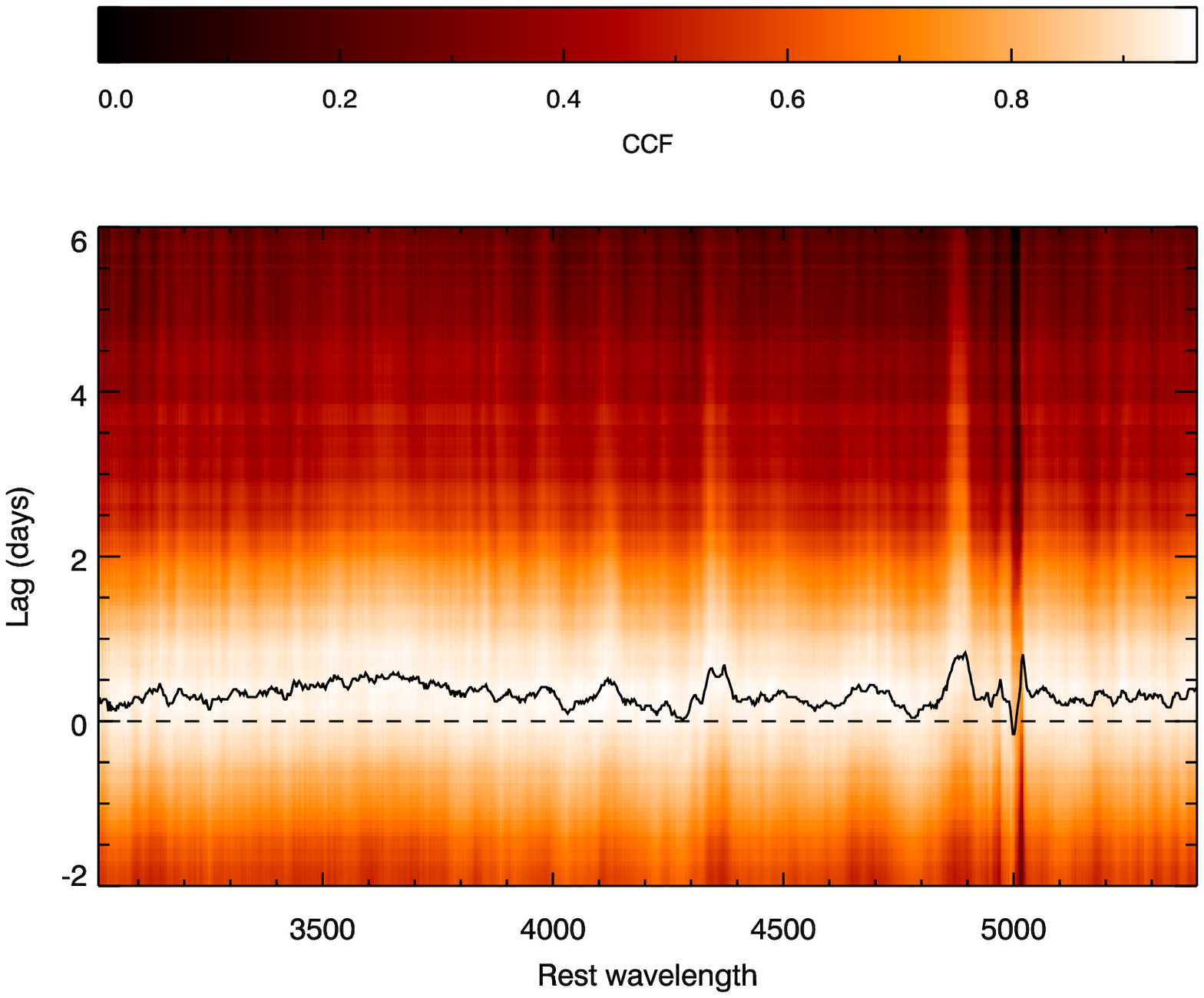}
\caption{Color map shows the CCF (with respect to {\it Swift}/W2) for the G430L grating.  The black line indicates the centroid of the CCF.}
\label{fig:ccfg430}
\end{figure}

\begin{figure}
\includegraphics[width=\columnwidth]{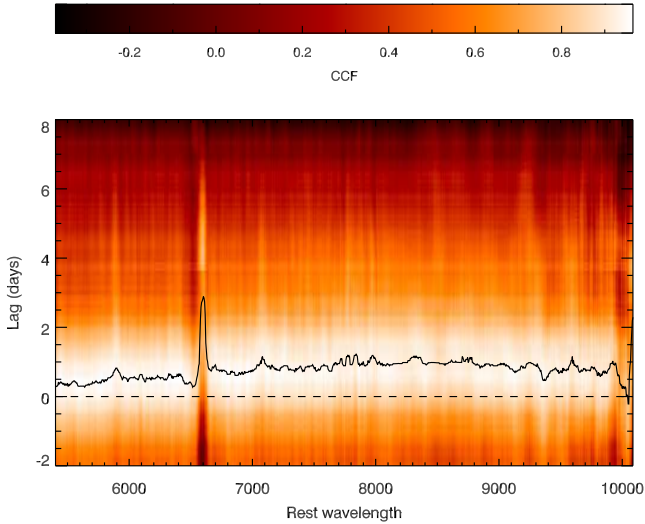}
\caption{Color map shows the CCF (with respect to {\it Swift}/W2) for the G750L grating.  The black line indicates the centroid of the CCF.}
\label{fig:ccfg750}
\end{figure}

To do this, for each wavelength bin in each spectrum we determine the mean flux within a small wavelength range (20\AA\ for G140L, 10\AA\ for the G430L and 30\AA\ for the G750L below 8000\AA\ and 100\AA\ above 8000\AA).  We slide a box of these widths across the whole spectrum in order to create lightcurves for every wavelength bin.  Each lightcurve is not independent of the neighboring lightcurve because of taking the mean over a few wavelength bins.  We then calculate the CCF with respect to the {\it Swift}/W2 lightcurve, and take the centroid of the CCF as the lag.  The lag spectrum for each grating is shown in panel (c) of Figs  \ref{fig:meanrmsg140}, \ref{fig:meanrmsg430}, \ref{fig:meanrmsg750}, and \ref{fig:meanrms_all}, furthermore, we show the lag spectrum on its own (with uncertainties, described below) in Figure~\ref{fig:lagspec}.  We also plot the CCF at each wavelength as a color map in Figs \ref{fig:ccfg140}, \ref{fig:ccfg430}, \ref{fig:ccfg750}.  The solid line in each of those figures indicates the centroid of the CCF.  

As we saw with the analysis of individual continuum bands, the lag spectra and CCFs also show a general increase in the continuum lag with wavelength except around the Balmer jump, where a decrease in the continuum lag is seen.  Around the Paschen jump the lags also flatten and start to decrease.  Lags from the emission lines  are clearly apparent in the lag spectra and CCFs, however, the emission line lags are not the focus of this work and will be addressed in a future paper.  We also perform FR only uncertainty calculations for the lag spectrum.  Since there are approximately $10^3$ wavelength bins for each of the three gratings it was computationally prohibitive to also run $10^5$ simulations in this case, and thus we only run $10^3$ simulations for each wavelength bin.  Note that since the sliding box goes from a width of 20\AA\ for the G430L to 30\AA\ for the G750L there is a sudden change in the size of the uncertainties at the boundary.

We also calculate the value of the correlation coefficient at the lag centroid ($r_{\rm cent}$), and plot this as a function of wavelength in panel (d) of Figs  \ref{fig:meanrmsg140}, \ref{fig:meanrmsg430}, \ref{fig:meanrmsg750} and \ref{fig:meanrms_all}.  This shows a very high correlation ($r_{\rm cent} > 0.9$) almost everywhere.  The few exceptions are (a) around the artifacts caused by the airglow lines at $\sim$1210\AA\ (this region of the spectrum should be ignored); (b) around the narrow [\ion{O}{3}] lines (4963, 5008\AA) -- those lines are not variable and the small residuals in the rms spectrum are artifacts of small wavelength shifts not yet calibrated out. This has no significant impact on the continuum analysis reported in this paper; (c) at the H$\alpha$ line (6563\AA).  There are other interesting wiggles in the correlation coefficient elsewhere, for instance around the absorption lines in \ion{C}{4}.  The drop in correlation at the H$\alpha$ line seems to be caused by a single anomalously low flux point in the lightcurves around that wavelength. Removal of that one point causes the correlation to go above 0.9 at H$\alpha$ while the continuum is not significantly affected. Since the emission lines are not part of this study we do not explore this further here.  

\section{Modeling the continuum lags}\label{sec:model}

The wavelength-dependent continuum lags show an excess and discontinuity around the Balmer jump.  This can clearly be seen when fitting the standard $\tau \propto \lambda^{4/3}$ relation to the observed lags (see solid line in Figure~\ref{fig:contlags}).  \citet{koristagoad01} discuss how the high-density clouds in the BLR should be efficient emitters of diffuse thermal continua, dominated by H free-bound emission and spanning the UV to near-IR.  This diffuse continuum emission should be present in both the flux spectrum and the measured continuum-band lags.  Here, we take an empirical approach to fitting the wavelength-dependent lags.  We leave a significantly more detailed and physically-motivated approach to account for diffuse continuum contribution to the measured lags over the full UV to near-IR spectrum to future work.

\subsection{Simple Model}

It is clear from the wavelength-dependence of the lags that there is a deviation around the Balmer jump.  While an enhanced lag in the $u$ and $U$ bands has been seen before in NGC~5548 \citep{edelson15,fausnaugh16} and NGC~4151 \citep{edelson17}, here with the spectroscopic coverage we are able to cover the wavelength region around the Balmer jump with many more continuum lightcurves, and resolve the feature there.  The largest deviations from the general trend with wavelength occur in the region from 3000 -- 4000 \AA.  Our first approach to modeling the lags is to ignore all lag measurements in that wavelength range and fit the $\tau \propto \lambda^{4/3}$ relation to the remaining points.  Since we perform the same uncertainty calculations on the W2 band (using the W2 as the both the reference and lightcurve of interest), we have an uncertainty measure on the zero-point. We therefore include the W2 point in the fit, and do not force the relation to have zero lag at the W2 wavelength.  In other words, we fit $\tau=\left(\tau_0\left(\lambda/\lambda_0\right)^{4/3}- 1\right)$ with $\tau_0$ and $\lambda_0$ as free parameters.  $\lambda_0$ is the reference wavelength where the lag vanishes, and $\tau_0$ is the lag at $\lambda_0$ relative to $\lambda=0$.  Our best-fit in this way gives $\tau_0 = 0.28 \pm 0.02$ days and $\lambda_0 = 2759 \pm 92$ \AA, and is shown as the solid line in Figure~\ref{fig:contlags}.  Uncertainties quoted on parameters are 1$\sigma$ confidence levels using the FR-only uncertainties on the data.   In the bottom panel of that Figure we show the residuals (data minus model).   This fit has a very poor reduced $\chi^2/\nu = 5.1$, for $\nu = 20$, driven by both the X-ray lag and under-predicting the lags between 2000 -- 3000 \AA\ (remember that 3000 -- 4000 \AA\ is excluded in the fit).  Despite the poor fit statistic,  it can be seen that the model fits the general trend well, with two exceptions (a) it clearly misses the X-ray lag by $0.40\pm0.05$~d \citep[the X-ray `offset', see ][for a further in depth discussion of this]{mchardy17}, and (b) the lags from $\sim$3000\AA\ to $\sim$4000\AA\ consistently lie above the model.  Thus, aside from the X-ray offset and the Balmer jump region, the lags broadly follow the basic $\tau \propto \lambda^{4/3}$ model.

We also try fitting a model with $\tau \propto \lambda^{a}$, with $a$ as a free parameter.  When including the X-ray lag, the slope of the fit is driven by this one point.  We therefore remove the X-ray lag to fit this model, finding $a = 1.45\pm0.17$, $\tau_0 = 0.20 \pm 0.04$ and $\lambda_0 = 2534\pm129$ \AA, with $\chi^2/\nu = 2.2$ for $\nu = 18$.  The resulting value of $a$ is consistent with $4/3$ within 1-$\sigma$.

The magnitude of the accretion disk lags relates to the size of the accretion disk.  We can compare the observed disk lags with those expected from the standard disk model based on NGC 4593's mass, and observed luminosity.  A convenient characterization of the lags expected from the standard disk model is given by Equation 12 from \citet{fausnaugh16}, which gives:
\begin{equation}
\tau_0 = \frac{1}{c} \left(X\frac{k\lambda_0}{hc} \right)^{4/3} \left[ \left(\frac{GM}{8\pi \sigma}\right)\left(\frac{L_{\rm Edd}}{\eta c^2}\right)\left(3 + \kappa\right)\,\dot{m}_{\rm E}\right]^{1/3}\ ,
\end{equation}
where we assume the ratio of external to internal heating, $\kappa = 1$ and that the radiative efficiency $\eta = 0.1$.   A range of radii contribute to emission at a given wavelength, and so to convert from temperature to wavelength for a given radius we assume a flux-weighted mean radius, which gives $X = 2.49$.  

To calculate the expected lags from the standard disk model we assume a mass of $M = 7.63\times10^6~M_\odot$ (we find the mass from \citealt{bentzkatz15}, which uses the H$\beta$ lags from \citealt{denney06} and \citealt{barth13}, assuming a virial factor of $f = 4.3$ from \citealt{grier13}).  We note that there is much uncertainty in determining bolometric corrections in order to estimate the Eddington fraction ($\dot{m}_{\rm E}$). The standard $L_{\rm bol} \simeq 9 \lambda L_\lambda$(5100 \AA) that we estimate from the mean spectrum gives $L_{\rm bol} = 2.1\times10^{43}$~erg~s$^{-1}$, which corresponds to $\dot{m}_{\rm E} = 0.022$.  This $L_{\rm bol}$ estimate is less than the X-ray luminosity based on {\it Swift} BAT measurements \citep[$L_{X} = 3\times10^{43}$~erg~s$^{-1}$,][]{mchardy17}.  We therefore use  $\dot{m}_{\rm E} = 0.081$ as estimated by \citet{mchardy17}, based on the spectral energy distribution fitting performed by \citet{vasudevan09,vasudevan10}.  With these parameters and $\lambda_0 = 2759$\AA\ we find 
$\tau_0 = 0.09$ days.  Our measured value of $\tau_0 = 0.28 \pm 0.02$ days is therefore a factor of 3.1 larger than expected from the standard disk model.

Here we have assumed the flux-weighted value for X.  If, however, one assumes that all the emission at a given wavelength comes from the annulus at the corresponding temperature given by Wien's law then X would be 4.97, leading to a model prediction a factor of $(4.97/2.49)^{4/3} = 2.5$ larger, and therefore consistent with the observed lags.   While this would be an unrealistic assumption, it demonstrates how different assumptions about X can affect the implied disk size.

\subsection{Contributions from the BLR}
The simple modeling approach above shows that there is a strong Balmer jump present in the lag spectrum.  Such a feature is expected from the model of \citet{koristagoad01}, who presented UV-to-near-IR flux and lag spectra of the diffuse continuum emitted by BLR clouds, as predicted from the same photoionization model that broadly reproduces the average luminosities and variability behaviors of the stronger UV broad emission lines in NGC 5548 during the 1989 and 1993 HST campaigns \citep{koristagoad00}.  This feature is also apparent at much lower resolution in the {\it Swift}-only monitoring of NGC~5548 \citep{edelson15}, NGC~4151 \citep{edelson17} and of this target, NGC~4593 \citep{mchardy17}.
In addition to the Balmer jump feature, \citet{koristagoad01} show that there should also be contributions from the diffuse continuum at all other wavelengths (UV to near-IR), with a generally rising contribution in flux toward longer wavelengths \citep{koristagoad01}.  In order to test for the presence of diffuse continuum lags from the BLR outside of the Balmer jump region requires detailed lightcurve simulations \citep[see e.g. the discussion in][]{fausnaugh16} and spectral decomposition which we leave to future work.  Here, we simply note that the significant discontinuity observed here in NGC 4593 is qualitatively consistent with the picture presented in \citet{koristagoad01}, where there is a significant contribution from the diffuse continuum in the BLR. 

\section{Discussion and Conclusion}\label{sec:disc}

Continuum reverberation mapping offers the opportunity to determine the temperature profile and size of the accretion disk through measuring wavelength-dependent time lags.  Here, we present 26 near daily {\it HST}/STIS observations of NGC~4593 in July and August 2016 that were taken as part of a multiwavelength campaign also involving {\it Swift} \citep[see][]{mchardy17}, {\it Kepler} and ground-based observations.  The STIS spectroscopy uses the low-resolution gratings (G140L, G430L and G750L) in order to cover as broad a wavelength range as possible, approximately 1150 -- 10000\AA\, with a 1700 -- 3000\AA\ gap that is nicely filled by {\it Swift's} UV bands.  One advantage of performing continuum reverberation with spectroscopy is that continuum and emission lines can be easily separated, which is not so straightforward when using broadband filters \citep[e.g.][]{chelouche13}.  Another advantage of our approach is the wavelength coverage around the Balmer jump, where lags arising from diffuse continuum emission in the BLR will be most prominent \citep{koristagoad01}.  

We find significant wavelength-dependent lags, with the continuum at 1150\AA\ leading the variations at 8950\AA\ by approximately 1.2 days.  The lags increase monotonically with wavelength except around the Balmer jump.  There, a clear discontinuity is observed, with the lags increasing to a peak at $\sim$3700\AA\ and then dropping to a local minimum at around 4200\AA.  For instance, the 3700\AA\ lightcurve lags the 4200\AA\ lightcurve by $0.30 \pm 0.07$ days.   A simple $\tau \propto \lambda^{4/3}$ relation therefore does not match the lags well between 3000--4000\AA.

The strongest previous evidence for a significant contribution to the continuum lags from diffuse continuum from BLR gas comes from the recent studies of NGC~5548 \citep{edelson15, fausnaugh16}.  In those studies, the lag from the $u$ and $U$ band filters were seen to deviate above the general wavelength-dependence from the other filters.  Here, however, since we have lags from spectroscopy, we can resolve the shape of the lags around the Balmer jump.

To model the lags we take an empirical approach by omitting the lags between 3000 -- 4000\AA\ from the fit.  Doing this, we find that $\tau \propto \lambda^{4/3}$ matches the general shape well outside of the 3000 -- 4000\AA\ region, though the X-ray lag is offset by $0.40\pm0.05$~d from the model.  The size of the accretion disk implied from this fit is about 3 times larger than expected from the standard disk  model.  However, if we accept that the lags at the Balmer jump are caused by diffuse continuum from the BLR then there should be diffuse continuum lags at all wavelengths.  While applying such a model to these data is beyond the scope of this paper, we note that the enhanced lags around the Balmer jump are qualitatively consistent with the model of \citet{koristagoad01}.

Recent 3D magnetohydrodynamic (MHD) simulations have highlighted the importance of the iron opacity bump in AGN accretion disks \citep{jiang16}. This can change the structure of the accretion disk at mid-plane temperatures of $\sim$$2\times10^5$~K.  Since it changes the disk thermal properties, it could change the wavelength-dependent lags \citep{jiang17}.  Temperature profiles and disk heights from such MHD simulations could be compared to the observed lags in NGC~4593 to test this.

Several recent studies of continuum reverberation have come to the conclusion that the accretion disks are larger than predicted by the standard thin disk model (see the many references in Section~\ref{sec:intro}).  Here, we are able to cleanly separate lags from the emission lines and continuum, and to estimate the contribution of the diffuse broad line gas to the continuum lags.  Even after taking this into account, we still find that the disk lags are a factor of $\sim$3 larger than the standard thin disk model, and the discrepancy remains.  This is similar in size to the discrepancy observed in other sources, though those typically do not fully account for the diffuse continuum lags. Microlensing studies have also concluded that accretion disks appear to be larger than the standard model \citep{morgan10,mosquera13}.  Significant contribution to the continuum light from the diffuse continuum in the broad line region needs to be taken into account there also and should act to decrease the discrepancy between the standard model and observations.  

The role of the X-rays in driving the variability at longer wavelengths has long been an open question \citep[e.g.,][]{uttley03, breedt09}.  Recent intensive {\it Swift} campaigns show that this issue remains unclear.  While some objects show excellent correlation between the X-ray and UV/optical lightcurves \citep{shappee14,troyer16}, the relationship between the X-ray and UV/optical lightcurves of NGC~5548 from the AGN STORM campaign is not straightforward.   \citet{starkey17} show that for a disk reprocessing model the driving lightcurve needed to fit the UV and optical variations does not resemble a blurred version of the X-ray lightcurve. \citet{gardnerdone17} argue that the observed X-ray lightcurve cannot be the driver, based on energetic considerations as well as on the lightcurve shapes, and invoke a 2-stage reprocessing model in which X-rays heat a vertically-extended torus producing EUV radiation that subsequently irradiates the disk. NGC~4151 shows a good X-ray/UV/optical correlation, but, also shows a much larger X-ray to UV lag than expected based on extrapolating the UV/optical lags \citep{edelson17}.  When looking at only the {\it Swift} data for NGC~4593 from our campaign, \citet{mchardy17} note that there is also a significant X-ray offset in this object, based on extrapolating the UV/optical lags.  When including the additional wavelength coverage from {\it HST} and fitting a $\lambda^{4/3}$ relation (ignoring the Balmer jump region) we also find a significant X-ray offset of $\sim$ 0.4 days.  Such an offset is at odds with the standard disk model.

We note that \citet{mchardy17} find a somewhat flatter $\tau(\lambda)$ relation from CCF analysis of {\it Swift} lightcurves covering 2200 -- 5400\AA, fully consistent with the $\tau\propto\lambda^{4/3}$ law for disk reprocessing, but with a highly significant X-ray lag offset of 0.7~d. The {\it HST} data extending {\it Swift's} coverage to longer and shorter wavelengths gives a steeper $\tau(\lambda)$ trend, and resolves the Balmer jump to more clearly discern
the diffuse continuum contribution to the CCF lags. \citet{mchardy17} were able to fit the {\it Swift} lightcurves by convolving the X-ray lightcurve with delay maps that feature a sharp component inside 2d, and a broad delay tail extending to 10d, interpretable as disk and BLR reprocessing, respectively.  That picture is consistent with the strong Balmer jump lags we observe.  

In conclusion, our spectroscopic monitoring of NGC 4593 with {\it HST} has demonstrated that the diffuse continuum from the BLR gas makes a significant contribution to the continuum time lags.  Similar future campaigns will help understand the contribution of this diffuse continuum in other AGN, which in turn leads to a better understanding of accretion disk sizes.

\acknowledgements
We thank the {\it Hubble} and \swi\ teams for their hard work and efforts in successfully scheduling this monitoring campaign.   Support for program number 14121 was provided by NASA through a grant from the Space Telescope Science Institute, which is operated by the Association of Universities for Research in Astronomy, Incorporated, under NASA contract NAS5-26555.  KH acknowledges support from STFC grant ST/M001296/1.  EMC thanks Kayhan G\"{u}ltekin for helpful discussions on the 2D lag centroid distributions.

\bibliographystyle{apj}
\bibliography{apj-jour,agn}

\appendix
\input{lightcurves_g140.tex}

\input{lightcurves_g430.tex}

\input{lightcurves_g750.tex}

\end{document}

%% file: lightcurves_g140.tex
\begin{deluxetable*}{ccccccc}[hb]
\tablecaption{{\it HST}/STIS continuum lightcurves from the G140L. \label{tab:lcs_g140}
}
\tablehead{
MJD - 57000 & \multicolumn{6}{c}{Flux density ($10^{-14}$ erg s$^{-1}$ cm$^{-2}$ \AA$^{-1}$)} \\
  & 1150\AA & 1310\AA & 1350\AA & 1460\AA & 1610\AA & 1690\AA }
  \startdata
581.662  & $12.38\pm 0.26$ & $14.00\pm 0.21$ & $12.25\pm 0.10$ & $11.64\pm 0.14$ & $15.86\pm 0.26$ & $13.34\pm 0.36 $ \\
582.656  & $12.91\pm 0.30$ & $14.78\pm 0.21$ & $13.25\pm 0.14$ & $12.43\pm 0.16$ & $17.09\pm 0.26$ & $13.68\pm 0.33 $ \\
583.650  & $ 9.85\pm 0.23$ & $11.56\pm 0.21$ & $10.05\pm 0.12$ & $ 9.27\pm 0.13$ & $13.65\pm 0.24$ & $10.59\pm 0.33 $ \\
583.782  & $ 9.75\pm 0.29$ & $11.33\pm 0.15$ & $10.25\pm 0.12$ & $ 9.38\pm 0.11$ & $13.86\pm 0.20$ & $11.08\pm 0.38 $ \\
584.578  & $ 8.30\pm 0.24$ & $10.14\pm 0.20$ & $ 8.93\pm 0.15$ & $ 8.02\pm 0.12$ & $12.46\pm 0.21$ & $ 9.62\pm 0.26 $ \\
585.571  & $ 6.91\pm 0.27$ & $ 8.85\pm 0.18$ & $ 7.63\pm 0.09$ & $ 7.04\pm 0.10$ & $10.85\pm 0.23$ & $ 8.21\pm 0.32 $ \\
586.542  & $ 7.73\pm 0.25$ & $ 9.75\pm 0.23$ & $ 8.09\pm 0.15$ & $ 7.59\pm 0.11$ & $10.96\pm 0.29$ & $ 8.28\pm 0.35 $ \\
587.559  & $12.10\pm 0.28$ & $13.85\pm 0.20$ & $11.98\pm 0.15$ & $10.99\pm 0.13$ & $15.12\pm 0.26$ & $12.20\pm 0.28 $ \\
588.486  & $13.51\pm 0.29$ & $15.01\pm 0.20$ & $13.22\pm 0.22$ & $12.12\pm 0.10$ & $16.70\pm 0.25$ & $13.60\pm 0.27 $ \\
589.414  & $13.95\pm 0.31$ & $14.65\pm 0.22$ & $13.60\pm 0.19$ & $12.25\pm 0.14$ & $17.28\pm 0.27$ & $13.87\pm 0.24 $ \\
590.407  & $14.82\pm 0.37$ & $15.65\pm 0.21$ & $14.54\pm 0.13$ & $12.92\pm 0.16$ & $17.31\pm 0.26$ & $13.71\pm 0.32 $ \\
591.401  & $11.14\pm 0.34$ & $12.91\pm 0.19$ & $11.47\pm 0.15$ & $10.48\pm 0.12$ & $15.25\pm 0.27$ & $11.61\pm 0.41 $ \\
592.394  & $ 9.71\pm 0.20$ & $11.65\pm 0.19$ & $10.25\pm 0.18$ & $ 9.36\pm 0.13$ & $13.86\pm 0.22$ & $10.76\pm 0.39 $ \\
593.388  & $10.04\pm 0.25$ & $11.79\pm 0.21$ & $10.50\pm 0.11$ & $ 9.47\pm 0.14$ & $13.17\pm 0.25$ & $11.09\pm 0.34 $ \\
594.315  & $11.08\pm 0.28$ & $12.22\pm 0.19$ & $10.94\pm 0.14$ & $10.02\pm 0.08$ & $14.26\pm 0.29$ & $10.57\pm 0.30 $ \\
595.309  & $10.28\pm 0.28$ & $11.93\pm 0.16$ & $10.62\pm 0.12$ & $ 9.78\pm 0.13$ & $14.17\pm 0.25$ & $11.14\pm 0.31 $ \\
596.428  & $ 8.90\pm 0.24$ & $10.29\pm 0.19$ & $ 9.24\pm 0.14$ & $ 8.13\pm 0.14$ & $12.35\pm 0.19$ & $ 9.24\pm 0.30 $ \\
597.422  & $ 8.12\pm 0.24$ & $ 9.55\pm 0.19$ & $ 8.43\pm 0.16$ & $ 7.61\pm 0.09$ & $11.37\pm 0.21$ & $ 9.07\pm 0.27 $ \\
598.289  & $ 7.16\pm 0.21$ & $ 9.00\pm 0.19$ & $ 7.59\pm 0.13$ & $ 6.93\pm 0.12$ & $ 9.90\pm 0.19$ & $ 8.09\pm 0.30 $ \\
599.283  & $ 8.71\pm 0.22$ & $ 9.86\pm 0.17$ & $ 8.45\pm 0.10$ & $ 7.81\pm 0.09$ & $10.97\pm 0.20$ & $ 9.52\pm 0.31 $ \\
600.210  & $ 6.21\pm 0.18$ & $ 8.07\pm 0.18$ & $ 6.93\pm 0.11$ & $ 6.13\pm 0.10$ & $ 8.92\pm 0.22$ & $ 7.37\pm 0.36 $ \\
602.197  & $ 6.96\pm 0.20$ & $ 8.34\pm 0.19$ & $ 7.09\pm 0.08$ & $ 6.14\pm 0.10$ & $ 7.92\pm 0.16$ & $ 7.32\pm 0.35 $ \\
603.191  & $ 5.67\pm 0.21$ & $ 7.48\pm 0.18$ & $ 6.29\pm 0.13$ & $ 5.62\pm 0.10$ & $ 7.93\pm 0.15$ & $ 6.11\pm 0.26 $ \\
604.185  & $ 6.31\pm 0.19$ & $ 7.86\pm 0.17$ & $ 6.69\pm 0.07$ & $ 6.12\pm 0.10$ & $ 8.25\pm 0.18$ & $ 7.19\pm 0.30 $ \\
605.238  & $ 5.67\pm 0.16$ & $ 7.27\pm 0.15$ & $ 6.06\pm 0.10$ & $ 5.63\pm 0.09$ & $ 7.63\pm 0.15$ & $ 6.51\pm 0.27 $ \\
606.172  & $ 6.52\pm 0.17$ & $ 7.78\pm 0.15$ & $ 6.25\pm 0.11$ & $ 5.83\pm 0.10$ & $ 7.81\pm 0.18$ & $ 6.99\pm 0.30 $ 
 \enddata
\end{deluxetable*}

%% file: lightcurves_g430.tex
 \clearpage
 \begin{turnpage}
 \begin{deluxetable*}{cccccccccccc}
\tablecaption{{\it HST}/STIS continuum lightcurves from the G430L. \label{tab:lcs_g430}}
\tabletypesize{\scriptsize}
\tablehead{
\colhead{MJD - 57000} & \multicolumn{11}{c}{Flux density ($10^{-14}$ erg s$^{-1}$ cm$^{-2}$ \AA$^{-1}$)} \\
\colhead{}  & \colhead{3050\AA} & \colhead{3250\AA} & \colhead{3550\AA} & \colhead{3700\AA} & \colhead{3800\AA} & \colhead{3925\AA} & \colhead{4200\AA} & \colhead{4430\AA} & \colhead{4745\AA} & \colhead{5100\AA} & \colhead{5450\AA} }
\startdata
581.662  & $10.64\pm 0.12$ & $11.11\pm 0.14$ & $ 8.46\pm 0.12$ & $ 8.01\pm 0.08$ & $ 6.84\pm 0.05$ & $ 6.06\pm 0.08$ & $ 4.18\pm 0.02$ & $ 4.11\pm 0.04$ & $ 3.46\pm 0.03$ & $ 3.15\pm 0.04$ & $ 2.60\pm 0.03 $ \\
582.656  & $10.75\pm 0.15$ & $11.29\pm 0.15$ & $ 8.35\pm 0.13$ & $ 8.09\pm 0.06$ & $ 6.85\pm 0.09$ & $ 6.02\pm 0.08$ & $ 4.27\pm 0.03$ & $ 4.21\pm 0.03$ & $ 3.54\pm 0.05$ & $ 3.15\pm 0.04$ & $ 2.61\pm 0.03 $ \\
583.650  & $10.11\pm 0.20$ & $10.67\pm 0.12$ & $ 8.21\pm 0.10$ & $ 7.80\pm 0.06$ & $ 6.53\pm 0.08$ & $ 5.66\pm 0.08$ & $ 3.87\pm 0.03$ & $ 3.89\pm 0.02$ & $ 3.20\pm 0.04$ & $ 2.99\pm 0.05$ & $ 2.52\pm 0.03 $ \\
583.782  & $10.21\pm 0.11$ & $11.01\pm 0.12$ & $ 8.51\pm 0.15$ & $ 7.91\pm 0.07$ & $ 6.67\pm 0.08$ & $ 5.85\pm 0.10$ & $ 3.96\pm 0.03$ & $ 3.90\pm 0.03$ & $ 3.20\pm 0.03$ & $ 3.02\pm 0.05$ & $ 2.53\pm 0.03 $ \\
584.578  & $ 9.44\pm 0.12$ & $10.02\pm 0.11$ & $ 7.86\pm 0.10$ & $ 7.47\pm 0.06$ & $ 6.25\pm 0.08$ & $ 5.55\pm 0.08$ & $ 3.78\pm 0.02$ & $ 3.76\pm 0.04$ & $ 3.02\pm 0.03$ & $ 2.82\pm 0.04$ & $ 2.38\pm 0.02 $ \\
585.571  & $ 8.47\pm 0.16$ & $ 9.30\pm 0.11$ & $ 7.16\pm 0.08$ & $ 6.72\pm 0.06$ & $ 5.80\pm 0.08$ & $ 5.15\pm 0.09$ & $ 3.41\pm 0.02$ & $ 3.40\pm 0.03$ & $ 2.62\pm 0.02$ & $ 2.55\pm 0.04$ & $ 2.09\pm 0.02 $ \\
586.542  & $ 8.06\pm 0.21$ & $ 8.66\pm 0.10$ & $ 6.47\pm 0.07$ & $ 5.98\pm 0.04$ & $ 5.29\pm 0.05$ & $ 4.78\pm 0.08$ & $ 3.20\pm 0.02$ & $ 3.17\pm 0.04$ & $ 2.43\pm 0.02$ & $ 2.39\pm 0.04$ & $ 1.92\pm 0.02 $ \\
587.559  & $10.19\pm 0.14$ & $10.72\pm 0.07$ & $ 7.96\pm 0.07$ & $ 7.39\pm 0.10$ & $ 6.48\pm 0.08$ & $ 5.66\pm 0.08$ & $ 3.97\pm 0.03$ & $ 3.87\pm 0.04$ & $ 3.16\pm 0.02$ & $ 2.91\pm 0.04$ & $ 2.39\pm 0.03 $ \\
588.486  & $10.73\pm 0.18$ & $11.28\pm 0.13$ & $ 8.54\pm 0.11$ & $ 7.93\pm 0.07$ & $ 6.76\pm 0.10$ & $ 6.05\pm 0.08$ & $ 4.21\pm 0.02$ & $ 4.12\pm 0.03$ & $ 3.45\pm 0.03$ & $ 3.09\pm 0.04$ & $ 2.55\pm 0.03 $ \\
589.414  & $10.50\pm 0.20$ & $10.99\pm 0.11$ & $ 8.25\pm 0.11$ & $ 7.95\pm 0.07$ & $ 6.74\pm 0.09$ & $ 5.97\pm 0.07$ & $ 4.17\pm 0.02$ & $ 4.16\pm 0.04$ & $ 3.39\pm 0.03$ & $ 3.07\pm 0.04$ & $ 2.51\pm 0.03 $ \\
590.407  & $10.28\pm 0.16$ & $10.81\pm 0.08$ & $ 8.19\pm 0.08$ & $ 7.91\pm 0.08$ & $ 6.64\pm 0.08$ & $ 5.96\pm 0.09$ & $ 4.12\pm 0.03$ & $ 4.05\pm 0.04$ & $ 3.34\pm 0.04$ & $ 3.00\pm 0.04$ & $ 2.50\pm 0.03 $ \\
591.401  & $10.44\pm 0.10$ & $10.83\pm 0.10$ & $ 8.09\pm 0.08$ & $ 7.69\pm 0.05$ & $ 6.46\pm 0.06$ & $ 5.77\pm 0.10$ & $ 4.01\pm 0.03$ & $ 3.96\pm 0.02$ & $ 3.28\pm 0.05$ & $ 2.98\pm 0.04$ & $ 2.49\pm 0.03 $ \\
592.394  & $ 9.17\pm 0.17$ & $10.00\pm 0.09$ & $ 7.69\pm 0.14$ & $ 7.28\pm 0.09$ & $ 6.31\pm 0.08$ & $ 5.45\pm 0.06$ & $ 3.76\pm 0.02$ & $ 3.72\pm 0.04$ & $ 2.99\pm 0.03$ & $ 2.78\pm 0.04$ & $ 2.30\pm 0.03 $ \\
593.388  & $ 9.62\pm 0.14$ & $10.00\pm 0.14$ & $ 7.68\pm 0.07$ & $ 7.26\pm 0.08$ & $ 6.26\pm 0.09$ & $ 5.51\pm 0.07$ & $ 3.79\pm 0.02$ & $ 3.71\pm 0.02$ & $ 2.96\pm 0.04$ & $ 2.80\pm 0.04$ & $ 2.32\pm 0.03 $ \\
594.315  & $ 9.41\pm 0.14$ & $10.09\pm 0.10$ & $ 7.70\pm 0.15$ & $ 7.31\pm 0.08$ & $ 6.20\pm 0.11$ & $ 5.41\pm 0.09$ & $ 3.74\pm 0.02$ & $ 3.68\pm 0.03$ & $ 3.00\pm 0.03$ & $ 2.75\pm 0.05$ & $ 2.28\pm 0.02 $ \\
595.309  & $ 9.41\pm 0.15$ & $10.08\pm 0.14$ & $ 7.53\pm 0.16$ & $ 7.24\pm 0.08$ & $ 6.10\pm 0.05$ & $ 5.44\pm 0.05$ & $ 3.73\pm 0.02$ & $ 3.59\pm 0.03$ & $ 2.97\pm 0.04$ & $ 2.73\pm 0.04$ & $ 2.32\pm 0.02 $ \\
596.428  & $ 9.34\pm 0.12$ & $10.05\pm 0.12$ & $ 7.58\pm 0.11$ & $ 7.12\pm 0.07$ & $ 6.00\pm 0.08$ & $ 5.24\pm 0.08$ & $ 3.49\pm 0.02$ & $ 3.44\pm 0.03$ & $ 2.82\pm 0.03$ & $ 2.63\pm 0.04$ & $ 2.22\pm 0.02 $ \\
597.422  & $ 9.03\pm 0.16$ & $ 9.72\pm 0.10$ & $ 7.18\pm 0.11$ & $ 6.78\pm 0.07$ & $ 5.73\pm 0.08$ & $ 5.03\pm 0.07$ & $ 3.44\pm 0.03$ & $ 3.40\pm 0.03$ & $ 2.73\pm 0.03$ & $ 2.60\pm 0.04$ & $ 2.15\pm 0.03 $ \\
598.289  & $ 8.33\pm 0.14$ & $ 9.11\pm 0.10$ & $ 6.81\pm 0.11$ & $ 6.39\pm 0.08$ & $ 5.52\pm 0.09$ & $ 4.95\pm 0.08$ & $ 3.32\pm 0.02$ & $ 3.33\pm 0.05$ & $ 2.54\pm 0.03$ & $ 2.47\pm 0.04$ & $ 2.02\pm 0.03 $ \\
599.283  & $ 8.04\pm 0.14$ & $ 8.88\pm 0.09$ & $ 6.56\pm 0.06$ & $ 6.25\pm 0.08$ & $ 5.40\pm 0.08$ & $ 4.84\pm 0.06$ & $ 3.27\pm 0.02$ & $ 3.21\pm 0.04$ & $ 2.50\pm 0.02$ & $ 2.40\pm 0.04$ & $ 1.96\pm 0.02 $ \\
600.210  & $ 7.98\pm 0.18$ & $ 8.57\pm 0.10$ & $ 6.58\pm 0.10$ & $ 6.28\pm 0.05$ & $ 5.30\pm 0.08$ & $ 4.74\pm 0.07$ & $ 3.13\pm 0.03$ & $ 3.13\pm 0.04$ & $ 2.39\pm 0.03$ & $ 2.34\pm 0.04$ & $ 1.95\pm 0.02 $ \\
602.197  & $ 7.29\pm 0.16$ & $ 7.83\pm 0.10$ & $ 5.66\pm 0.09$ & $ 5.46\pm 0.07$ & $ 4.68\pm 0.07$ & $ 4.26\pm 0.06$ & $ 2.86\pm 0.02$ & $ 2.83\pm 0.05$ & $ 2.14\pm 0.03$ & $ 2.09\pm 0.05$ & $ 1.72\pm 0.03 $ \\
603.191  & $ 7.27\pm 0.13$ & $ 7.96\pm 0.08$ & $ 5.92\pm 0.06$ & $ 5.57\pm 0.06$ & $ 4.79\pm 0.07$ & $ 4.30\pm 0.05$ & $ 2.94\pm 0.02$ & $ 2.95\pm 0.03$ & $ 2.26\pm 0.03$ & $ 2.18\pm 0.04$ & $ 1.76\pm 0.03 $ \\
604.185  & $ 7.41\pm 0.17$ & $ 8.12\pm 0.09$ & $ 5.91\pm 0.07$ & $ 5.57\pm 0.07$ & $ 4.81\pm 0.07$ & $ 4.34\pm 0.04$ & $ 2.88\pm 0.02$ & $ 2.90\pm 0.04$ & $ 2.22\pm 0.03$ & $ 2.15\pm 0.04$ & $ 1.75\pm 0.03 $ \\
605.238  & $ 7.43\pm 0.22$ & $ 8.25\pm 0.08$ & $ 6.01\pm 0.14$ & $ 5.84\pm 0.07$ & $ 4.98\pm 0.07$ & $ 4.51\pm 0.06$ & $ 2.94\pm 0.02$ & $ 2.88\pm 0.03$ & $ 2.25\pm 0.03$ & $ 2.21\pm 0.04$ & $ 1.80\pm 0.02 $ \\
606.172  & $ 7.18\pm 0.16$ & $ 7.71\pm 0.08$ & $ 5.73\pm 0.09$ & $ 5.52\pm 0.07$ & $ 4.70\pm 0.08$ & $ 4.33\pm 0.06$ & $ 2.79\pm 0.02$ & $ 2.78\pm 0.04$ & $ 2.12\pm 0.02$ & $ 2.08\pm 0.04$ & $ 1.73\pm 0.02 $ 
\enddata
\end{deluxetable*}
\end{turnpage}
\clearpage

%% file: lightcurves_g750.tex
 \begin{deluxetable*}{cccccccc}
\tablecaption{{\it HST}/STIS continuum lightcurves from the G750L. \label{tab:lcs_g750}}
\tablehead{
MJD - 57000 & \multicolumn{7}{c}{Flux density ($10^{-14}$ erg s$^{-1}$ cm$^{-2}$ \AA$^{-1}$)} \\
  & 5600\AA & 6250\AA & 6850\AA & 7450\AA & 8000\AA & 8800\AA & 9350\AA }
\startdata
581.662  & $ 2.42\pm 0.02$ & $ 2.44\pm 0.01$ & $ 2.13\pm 0.02$ & $ 2.06\pm 0.01$ & $ 1.94\pm 0.02$ & $ 1.68\pm 0.01$ & $ 1.41\pm 0.05 $ \\
582.656  & $ 2.45\pm 0.03$ & $ 2.46\pm 0.03$ & $ 2.17\pm 0.01$ & $ 2.06\pm 0.02$ & $ 1.93\pm 0.02$ & $ 1.68\pm 0.02$ & $ 1.37\pm 0.05 $ \\
583.650  & $ 2.30\pm 0.02$ & $ 2.37\pm 0.02$ & $ 2.05\pm 0.01$ & $ 1.97\pm 0.01$ & $ 1.85\pm 0.02$ & $ 1.59\pm 0.02$ & $ 1.34\pm 0.05 $ \\
583.782  & $ 2.35\pm 0.02$ & $ 2.39\pm 0.02$ & $ 2.07\pm 0.01$ & $ 1.97\pm 0.01$ & $ 1.85\pm 0.02$ & $ 1.58\pm 0.02$ & $ 1.32\pm 0.04 $ \\
584.578  & $ 2.20\pm 0.02$ & $ 2.27\pm 0.02$ & $ 2.04\pm 0.02$ & $ 1.98\pm 0.01$ & $ 1.87\pm 0.02$ & $ 1.63\pm 0.03$ & $ 1.34\pm 0.04 $ \\
585.571  & $ 1.97\pm 0.05$ & $ 2.10\pm 0.02$ & $ 1.87\pm 0.02$ & $ 1.85\pm 0.02$ & $ 1.76\pm 0.02$ & $ 1.55\pm 0.03$ & $ 1.24\pm 0.06 $ \\
586.542  & $ 1.81\pm 0.03$ & $ 1.82\pm 0.02$ & $ 1.65\pm 0.01$ & $ 1.62\pm 0.01$ & $ 1.54\pm 0.02$ & $ 1.36\pm 0.02$ & $ 1.19\pm 0.04 $ \\
587.559  & $ 2.24\pm 0.02$ & $ 2.26\pm 0.02$ & $ 1.93\pm 0.01$ & $ 1.85\pm 0.01$ & $ 1.72\pm 0.02$ & $ 1.51\pm 0.02$ & $ 1.31\pm 0.04 $ \\
588.486  & $ 2.42\pm 0.02$ & $ 2.41\pm 0.02$ & $ 2.05\pm 0.01$ & $ 1.92\pm 0.02$ & $ 1.80\pm 0.02$ & $ 1.57\pm 0.03$ & $ 1.32\pm 0.05 $ \\
589.414  & $ 2.36\pm 0.02$ & $ 2.37\pm 0.02$ & $ 2.07\pm 0.01$ & $ 1.96\pm 0.01$ & $ 1.85\pm 0.02$ & $ 1.63\pm 0.02$ & $ 1.36\pm 0.04 $ \\
590.407  & $ 2.33\pm 0.03$ & $ 2.34\pm 0.02$ & $ 2.07\pm 0.01$ & $ 2.00\pm 0.02$ & $ 1.88\pm 0.02$ & $ 1.62\pm 0.02$ & $ 1.38\pm 0.04 $ \\
591.401  & $ 2.32\pm 0.03$ & $ 2.34\pm 0.02$ & $ 2.05\pm 0.01$ & $ 1.95\pm 0.01$ & $ 1.84\pm 0.02$ & $ 1.63\pm 0.02$ & $ 1.37\pm 0.05 $ \\
592.394  & $ 2.17\pm 0.02$ & $ 2.23\pm 0.02$ & $ 2.00\pm 0.01$ & $ 1.92\pm 0.02$ & $ 1.85\pm 0.02$ & $ 1.62\pm 0.02$ & $ 1.36\pm 0.05 $ \\
593.388  & $ 2.12\pm 0.04$ & $ 2.21\pm 0.02$ & $ 1.95\pm 0.01$ & $ 1.90\pm 0.01$ & $ 1.83\pm 0.02$ & $ 1.57\pm 0.03$ & $ 1.34\pm 0.04 $ \\
594.315  & $ 2.20\pm 0.03$ & $ 2.24\pm 0.02$ & $ 1.98\pm 0.01$ & $ 1.97\pm 0.01$ & $ 1.80\pm 0.02$ & $ 1.59\pm 0.02$ & $ 1.30\pm 0.05 $ \\
595.309  & $ 2.18\pm 0.02$ & $ 2.23\pm 0.02$ & $ 1.98\pm 0.01$ & $ 1.96\pm 0.02$ & $ 1.83\pm 0.02$ & $ 1.59\pm 0.03$ & $ 1.31\pm 0.04 $ \\
596.428  & $ 2.04\pm 0.02$ & $ 2.10\pm 0.02$ & $ 1.81\pm 0.01$ & $ 1.74\pm 0.02$ & $ 1.66\pm 0.02$ & $ 1.44\pm 0.02$ & $ 1.25\pm 0.04 $ \\
597.422  & $ 2.02\pm 0.02$ & $ 2.08\pm 0.02$ & $ 1.81\pm 0.01$ & $ 1.73\pm 0.01$ & $ 1.65\pm 0.02$ & $ 1.49\pm 0.02$ & $ 1.21\pm 0.04 $ \\
598.289  & $ 1.90\pm 0.02$ & $ 1.99\pm 0.03$ & $ 1.77\pm 0.01$ & $ 1.78\pm 0.02$ & $ 1.68\pm 0.02$ & $ 1.48\pm 0.02$ & $ 1.24\pm 0.04 $ \\
599.283  & $ 1.81\pm 0.02$ & $ 1.94\pm 0.02$ & $ 1.71\pm 0.01$ & $ 1.70\pm 0.02$ & $ 1.65\pm 0.02$ & $ 1.43\pm 0.02$ & $ 1.23\pm 0.04 $ \\
600.210  & $ 1.78\pm 0.03$ & $ 1.93\pm 0.02$ & $ 1.68\pm 0.02$ & $ 1.70\pm 0.01$ & $ 1.62\pm 0.02$ & $ 1.46\pm 0.02$ & $ 1.17\pm 0.04 $ \\
602.197  & $ 1.58\pm 0.02$ & $ 1.71\pm 0.02$ & $ 1.49\pm 0.01$ & $ 1.49\pm 0.02$ & $ 1.43\pm 0.01$ & $ 1.31\pm 0.03$ & $ 1.06\pm 0.04 $ \\
603.191  & $ 1.65\pm 0.02$ & $ 1.77\pm 0.02$ & $ 1.54\pm 0.00$ & $ 1.54\pm 0.01$ & $ 1.47\pm 0.02$ & $ 1.31\pm 0.03$ & $ 1.13\pm 0.04 $ \\
604.185  & $ 1.65\pm 0.02$ & $ 1.74\pm 0.02$ & $ 1.48\pm 0.01$ & $ 1.47\pm 0.01$ & $ 1.42\pm 0.01$ & $ 1.28\pm 0.02$ & $ 1.09\pm 0.03 $ \\
605.238  & $ 1.68\pm 0.02$ & $ 1.78\pm 0.02$ & $ 1.51\pm 0.01$ & $ 1.49\pm 0.01$ & $ 1.42\pm 0.01$ & $ 1.30\pm 0.02$ & $ 1.08\pm 0.03 $ \\
606.172  & $ 1.64\pm 0.02$ & $ 1.70\pm 0.02$ & $ 1.51\pm 0.01$ & $ 1.52\pm 0.01$ & $ 1.48\pm 0.02$ & $ 1.33\pm 0.02$ & $ 1.12\pm 0.04 $ 
\enddata
\end{deluxetable*}

%% file: ms.bbl
\begin{thebibliography}{56}
\expandafter\ifx\csname natexlab\endcsname\relax\def\natexlab#1{#1}\fi

\bibitem[{{Anderson} \& {Bedin}(2010)}]{andersonbedin10}
{Anderson}, J. \& {Bedin}, L.~R. 2010, \pasp, 122, 1035

\bibitem[{{Barth} {et~al.}(2013)}]{barth13}
{Barth}, A.~J. {et~al.} 2013, \apj, 769, 128

\bibitem[{{Bentz} \& {Katz}(2015)}]{bentzkatz15}
{Bentz}, M.~C. \& {Katz}, S. 2015, \pasp, 127, 67

\bibitem[{{Bentz} {et~al.}(2009)}]{bentz09}
{Bentz}, M.~C. {et~al.} 2009, \apj, 705, 199

\bibitem[{{Blandford} \& {McKee}(1982)}]{blandmckee82}
{Blandford}, R.~D. \& {McKee}, C.~F. 1982, \apj, 255, 419

\bibitem[{{Breedt} {et~al.}(2009){Breedt}, {Ar{\'e}valo}, {McHardy}, {Uttley},
  {Sergeev}, {Minezaki}, {Yoshii}, {Gaskell}, {Cackett}, {Horne}, \&
  {Koshida}}]{breedt09}
{Breedt}, E., {Ar{\'e}valo}, P., {McHardy}, I.~M., {Uttley}, P., {Sergeev},
  S.~G., {Minezaki}, T., {Yoshii}, Y., {Gaskell}, C.~M., {Cackett}, E.~M.,
  {Horne}, K., \& {Koshida}, S. 2009, \mnras, 394, 427

\bibitem[{{Brewer} {et~al.}(2011){Brewer}, {Treu}, {Pancoast}, {Barth},
  {Bennert}, {Bentz}, {Filippenko}, {Greene}, {Malkan}, \& {Woo}}]{brewer11}
{Brewer}, B.~J., {Treu}, T., {Pancoast}, A., {Barth}, A.~J., {Bennert}, V.~N.,
  {Bentz}, M.~C., {Filippenko}, A.~V., {Greene}, J.~E., {Malkan}, M.~A., \&
  {Woo}, J.-H. 2011, \apjl, 733, L33

\bibitem[{{Buisson} {et~al.}(2017){Buisson}, {Lohfink}, {Alston}, \&
  {Fabian}}]{buisson17}
{Buisson}, D.~J.~K., {Lohfink}, A.~M., {Alston}, W.~N., \& {Fabian}, A.~C.
  2017, \mnras, 464, 3194

\bibitem[{{Cackett} {et~al.}(2007){Cackett}, {Horne}, \& {Winkler}}]{cackett07}
{Cackett}, E.~M., {Horne}, K., \& {Winkler}, H. 2007, \mnras, 380, 669

\bibitem[{{Cardelli} {et~al.}(1989){Cardelli}, {Clayton}, \&
  {Mathis}}]{cardelli89}
{Cardelli}, J.~A., {Clayton}, G.~C., \& {Mathis}, J.~S. 1989, \apj, 345, 245

\bibitem[{{Chelouche}(2013)}]{chelouche13}
{Chelouche}, D. 2013, \apj, 772, 9

\bibitem[{{Collier} {et~al.}(1999){Collier}, {Horne}, {Wanders}, \&
  {Peterson}}]{collier99}
{Collier}, S., {Horne}, K., {Wanders}, I., \& {Peterson}, B.~M. 1999, \mnras,
  302, L24

\bibitem[{{Collier} {et~al.}(2001)}]{collier01}
{Collier}, S. {et~al.} 2001, \apj, 561, 146

\bibitem[{{Collier} {et~al.}(1998)}]{collier98}
{Collier}, S.~J. {et~al.} 1998, \apj, 500, 162

\bibitem[{{Dai} {et~al.}(2010){Dai}, {Kochanek}, {Chartas}, {Koz{\l}owski},
  {Morgan}, {Garmire}, \& {Agol}}]{dai10}
{Dai}, X., {Kochanek}, C.~S., {Chartas}, G., {Koz{\l}owski}, S., {Morgan},
  C.~W., {Garmire}, G., \& {Agol}, E. 2010, \apj, 709, 278

\bibitem[{{De Rosa} {et~al.}(2015){De Rosa}, {Peterson}, {Ely},
  {et~al.}}]{derosa15}
{De Rosa}, G., {Peterson}, B.~M., {Ely}, J., {et~al.} 2015, \apj, 806, 128

\bibitem[{{Denney} {et~al.}(2006)}]{denney06}
{Denney}, K.~D. {et~al.} 2006, \apj, 653, 152

\bibitem[{{Edelson} {et~al.}(2017){Edelson}, {Gelbord}, {Cackett},
  {et~al.}}]{edelson17}
{Edelson}, R., {Gelbord}, J., {Cackett}, E., {et~al.} 2017, \apj, 840, 41

\bibitem[{{Edelson} {et~al.}(2015){Edelson}, {Gelbord}, {Horne},
  {et~al.}}]{edelson15}
{Edelson}, R., {Gelbord}, J.~M., {Horne}, K., {et~al.} 2015, \apj, 806, 129

\bibitem[{{Fausnaugh} {et~al.}(2016){Fausnaugh}, {Denney}, {Barth},
  {et~al.}}]{fausnaugh16}
{Fausnaugh}, M.~M., {Denney}, K.~D., {Barth}, A.~J., {et~al.} 2016, \apj, 821,
  56

\bibitem[{{Gardner} \& {Done}(2017)}]{gardnerdone17}
{Gardner}, E. \& {Done}, C. 2017, \mnras, 470, 3591

\bibitem[{{Gliozzi} {et~al.}(2017){Gliozzi}, {Papadakis}, {Grupe}, {Brinkmann},
  \& {R{\"a}th}}]{gliozzi17}
{Gliozzi}, M., {Papadakis}, I.~E., {Grupe}, D., {Brinkmann}, W.~P., \&
  {R{\"a}th}, C. 2017, \mnras, 464, 3955

\bibitem[{{Goad} {et~al.}(2016){Goad}, {Korista}, {De Rosa}, {et~al.}}]{goad16}
{Goad}, M.~R., {Korista}, K.~T., {De Rosa}, G., {et~al.} 2016, \apj, 824, 11

\bibitem[{{Grier} {et~al.}(2013){Grier}, {Martini}, {Watson}, {Peterson},
  {Bentz}, {Dasyra}, {Dietrich}, {Ferrarese}, {Pogge}, \& {Zu}}]{grier13}
{Grier}, C.~J., {Martini}, P., {Watson}, L.~C., {Peterson}, B.~M., {Bentz},
  M.~C., {Dasyra}, K.~M., {Dietrich}, M., {Ferrarese}, L., {Pogge}, R.~W., \&
  {Zu}, Y. 2013, \apj, 773, 90

\bibitem[{{Horne} {et~al.}(2004){Horne}, {Peterson}, {Collier}, \&
  {Netzer}}]{hornepcn04}
{Horne}, K., {Peterson}, B.~M., {Collier}, S.~J., \& {Netzer}, H. 2004, \pasp,
  116, 465

\bibitem[{{Jiang} {et~al.}(2016){Jiang}, {Davis}, \& {Stone}}]{jiang16}
{Jiang}, Y.-F., {Davis}, S.~W., \& {Stone}, J.~M. 2016, \apj, 827, 10

\bibitem[{{Jiang} {et~al.}(2017){Jiang}, {Green}, {Greene}, {et~al.}}]{jiang17}
{Jiang}, Y.-F., {Green}, P.~J., {Greene}, J.~E., {et~al.} 2017, \apj, 836, 186

\bibitem[{{Kara} {et~al.}(2016){Kara}, {Alston}, {Fabian}, {Cackett}, {Uttley},
  {Reynolds}, \& {Zoghbi}}]{kara16}
{Kara}, E., {Alston}, W.~N., {Fabian}, A.~C., {Cackett}, E.~M., {Uttley}, P.,
  {Reynolds}, C.~S., \& {Zoghbi}, A. 2016, \mnras, 462, 511

\bibitem[{{Korista} \& {Goad}(2000)}]{koristagoad00}
{Korista}, K.~T. \& {Goad}, M.~R. 2000, \apj, 536, 284

\bibitem[{{Korista} \& {Goad}(2001)}]{koristagoad01}
---. 2001, \apj, 553, 695

\bibitem[{{Mathur} {et~al.}(2017){Mathur}, {Gupta}, {Page},
  {et~al.}}]{mathur17}
{Mathur}, S., {Gupta}, A., {Page}, K., {et~al.} 2017, \apj, 846, 55

\bibitem[{{McHardy} {et~al.}(2014){McHardy}, {Cameron}, {Dwelly}, {Connolly},
  {Lira}, {Emmanoulopoulos}, {Gelbord}, {Breedt}, {Arevalo}, \&
  {Uttley}}]{mchardy14}
{McHardy}, I.~M., {Cameron}, D.~T., {Dwelly}, T., {Connolly}, S., {Lira}, P.,
  {Emmanoulopoulos}, D., {Gelbord}, J., {Breedt}, E., {Arevalo}, P., \&
  {Uttley}, P. 2014, \mnras, 444, 1469

\bibitem[{{McHardy} {et~al.}(2017){McHardy}, {Connolly}, {Horne},
  {et~al.}}]{mchardy17}
{McHardy}, I.~M., {Connolly}, S.~D., {Horne}, K., {et~al.} 2017, \mnras,
  submitted, arXiv:1712.04852

\bibitem[{{Morgan} {et~al.}(2010){Morgan}, {Kochanek}, {Morgan}, \&
  {Falco}}]{morgan10}
{Morgan}, C.~W., {Kochanek}, C.~S., {Morgan}, N.~D., \& {Falco}, E.~E. 2010,
  \apj, 712, 1129

\bibitem[{{Mosquera} {et~al.}(2013){Mosquera}, {Kochanek}, {Chen}, {Dai},
  {Blackburne}, \& {Chartas}}]{mosquera13}
{Mosquera}, A.~M., {Kochanek}, C.~S., {Chen}, B., {Dai}, X., {Blackburne},
  J.~A., \& {Chartas}, G. 2013, \apj, 769, 53

\bibitem[{{Mudd} {et~al.}(2017){Mudd}, {Martini}, {Zu}, \& {The DES
  Collaboration}}]{mudd17}
{Mudd}, D., {Martini}, P., {Zu}, Y., \& {The DES Collaboration}. 2017,
  submitted to ApJ, arXiv:1711.11588

\bibitem[{{Noda} {et~al.}(2016){Noda}, {Minezaki}, {Watanabe},
  {et~al.}}]{noda16}
{Noda}, H., {Minezaki}, T., {Watanabe}, M., {et~al.} 2016, \apj, 828, 78

\bibitem[{{Pal} {et~al.}(2016){Pal}, {Dewangan}, {Connolly}, \&
  {Misra}}]{pal16}
{Pal}, M., {Dewangan}, G.~C., {Connolly}, S.~D., \& {Misra}, R. 2016, ArXiv
  e-prints

\bibitem[{{Pal} \& {Naik}(2017)}]{palnaik17}
{Pal}, M. \& {Naik}, S. 2017, MNRAS, in press, arXiv:1711.11194

\bibitem[{{Pancoast} {et~al.}(2014){Pancoast}, {Brewer}, {Treu}, {Park},
  {Barth}, {Bentz}, \& {Woo}}]{pancoast14}
{Pancoast}, A., {Brewer}, B.~J., {Treu}, T., {Park}, D., {Barth}, A.~J.,
  {Bentz}, M.~C., \& {Woo}, J.-H. 2014, \mnras, 445, 3073

\bibitem[{{Pei} {et~al.}(2017){Pei}, {Fausnaugh}, {Barth}, {et~al.}}]{pei17}
{Pei}, L., {Fausnaugh}, M.~M., {Barth}, A.~J., {et~al.} 2017, \apj, 837, 131

\bibitem[{{Peterson}(2014)}]{peterson14}
{Peterson}, B.~M. 2014, \ssr, 183, 253

\bibitem[{{Peterson} {et~al.}(2004)}]{petersonetal04}
{Peterson}, B.~M. {et~al.} 2004, \apj, 613, 682

\bibitem[{{Schlafly} \& {Finkbeiner}(2011)}]{schlafly11}
{Schlafly}, E.~F. \& {Finkbeiner}, D.~P. 2011, \apj, 737, 103

\bibitem[{{Sergeev} {et~al.}(2005){Sergeev}, {Doroshenko}, {Golubinskiy},
  {Merkulova}, \& {Sergeeva}}]{sergeev05}
{Sergeev}, S.~G., {Doroshenko}, V.~T., {Golubinskiy}, Y.~V., {Merkulova},
  N.~I., \& {Sergeeva}, E.~A. 2005, \apj, 622, 129

\bibitem[{{Shappee} {et~al.}(2014)}]{shappee14}
{Shappee}, B.~J. {et~al.} 2014, \apj, 788, 48

\bibitem[{{Starkey} {et~al.}(2017){Starkey}, {Horne}, {Fausnaugh},
  {et~al.}}]{starkey17}
{Starkey}, D., {Horne}, K., {Fausnaugh}, M.~M., {et~al.} 2017, \apj, 835, 65

\bibitem[{{Troyer} {et~al.}(2016){Troyer}, {Starkey}, {Cackett}, {Bentz},
  {Goad}, {Horne}, \& {Seals}}]{troyer16}
{Troyer}, J., {Starkey}, D., {Cackett}, E.~M., {Bentz}, M.~C., {Goad}, M.~R.,
  {Horne}, K., \& {Seals}, J.~E. 2016, \mnras, 456, 4040

\bibitem[{{Uttley} {et~al.}(2014){Uttley}, {Cackett}, {Fabian}, {Kara}, \&
  {Wilkins}}]{uttley14}
{Uttley}, P., {Cackett}, E.~M., {Fabian}, A.~C., {Kara}, E., \& {Wilkins},
  D.~R. 2014, \aapr, 22, 72

\bibitem[{{Uttley} {et~al.}(2003){Uttley}, {Edelson}, {McHardy}, {Peterson}, \&
  {Markowitz}}]{uttley03}
{Uttley}, P., {Edelson}, R., {McHardy}, I.~M., {Peterson}, B.~M., \&
  {Markowitz}, A. 2003, \apjl, 584, L53

\bibitem[{{Vasudevan} \& {Fabian}(2009)}]{vasudevan09}
{Vasudevan}, R.~V. \& {Fabian}, A.~C. 2009, \mnras, 392, 1124

\bibitem[{{Vasudevan} {et~al.}(2010){Vasudevan}, {Fabian}, {Gandhi}, {Winter},
  \& {Mushotzky}}]{vasudevan10}
{Vasudevan}, R.~V., {Fabian}, A.~C., {Gandhi}, P., {Winter}, L.~M., \&
  {Mushotzky}, R.~F. 2010, \mnras, 402, 1081

\bibitem[{{Vaughan} {et~al.}(2003){Vaughan}, {Edelson}, {Warwick}, \&
  {Uttley}}]{vaughan03}
{Vaughan}, S., {Edelson}, R., {Warwick}, R.~S., \& {Uttley}, P. 2003, \mnras,
  345, 1271

\bibitem[{{Wanders} {et~al.}(1997)}]{wanders97}
{Wanders}, I. {et~al.} 1997, \apjs, 113, 69

\bibitem[{{Welsh} \& {Horne}(1991)}]{welsh91}
{Welsh}, W.~F. \& {Horne}, K. 1991, \apj, 379, 586

\bibitem[{{White} \& {Peterson}(1994)}]{white94}
{White}, R.~J. \& {Peterson}, B.~M. 1994, \pasp, 106, 879

\end{thebibliography}
